\documentclass[final,3p,times]{elsarticle}

\usepackage{dcolumn}
\usepackage{bm}
\usepackage{float}
\usepackage{epsfig}
\usepackage{graphicx}
\usepackage{longtable} 
\usepackage{rotating}
\usepackage{amsmath}
\usepackage{amsfonts}
\usepackage{amssymb}
\usepackage{units}
\usepackage{multirow}

\newcommand{\sqrts}{\sqrt{s}}
\newcommand{\sqrtsnn}{\sqrt{s_{_{\ensuremath{\it{NN}}}}}}
\newcommand{\pp}{$p$-$p$}
\newcommand{\ppbar}{$p$-$\bar p$}

\newcommand{\Pom}{\mathbb{P}}

\newcommand{\phojet}{\textsc{phojet}}
\newcommand{\dpmjet}{\textsc{dpmjet}}
\newcommand{\epos}{\textsc{epos}}
\newcommand{\nexus}{\textsc{neXus}}
\newcommand{\qgsjet}{\textsc{qgsjet}} 
\newcommand{\sibyll}{\textsc{sibyll}}

\newcommand{\pythia}{\textsc{pythia}}
\newcommand{\herwig}{\textsc{herwig}}
\newcommand{\lhapdf}{\textsc{lhapdf}}

\newcommand{\dNdeta}{dN_{ch}/d\eta|_{\eta=0}}
\newcommand{\meanpt}{\ensuremath{\left< p_{\perp} \right>}}
\newcommand{\pT}{p_{\perp}}

\def\Lambdaqcd{\Lambda_{\ensuremath{\it QCD}}}
\def\mean#1{\ensuremath{\left<#1\right>}}
\def\ttt#1{\texttt{\small #1}}
\def\cO#1{{{\cal{O}}}\rm{\left(#1\right)}}

\newcommand{\comment}[1]{}

\journal{Astroparticle Physics}

\begin{document}

\begin{frontmatter}


\title{Constraints from the first LHC data on hadronic event
  \\ generators for ultra-high energy cosmic-ray physics}


\author{David d'Enterria}\address{ICREA \& ICC-UB, Univ. de Barcelona, 08028 Barcelona, Catalonia \\
CERN, PH Department, 1211 Geneva, Switzerland}
\author{Ralph Engel, Tanguy Pierog}\address{Karlsruhe Institut of Technology, Postfach 3640, 76021 Karlsruhe, Germany}
\author{Sergey Ostapchenko}\address{NTNU, Inst. for Fysikk, 7491 Trondheim, Norway \\
D.V. Skobeltsyn Inst. Nuc. Phys, Moscow State Univ., 119992 Moscow, Russia}
\author{Klaus Werner}\address{SUBATECH, 4 rue Alfred Kastler, BP 20722, 44307 Nantes Cedex 3, France\\}


\begin{abstract}
  The determination of the primary energy and mass of ultra-high-energy cosmic-rays (UHECR)
  generating extensive air-showers in the Earth's atmosphere, relies on the 
  detailed modeling of hadronic multiparticle production at center-of-mass (c.m.) collision 
  energies up to two orders of magnitude higher than those studied at particle colliders.
  The first Large Hadron Collider (LHC) data have extended by more than a factor of three the c.m.~energies in which 
  we have direct proton-proton measurements available to compare to hadronic models. 
  In this work we compare LHC results on inclusive particle production at
  energies $\sqrts$ = 0.9, 2.36, and 7 TeV to predictions of various
  hadronic Monte Carlo (MC) models used commonly in cosmic-ray (CR) physics (\qgsjet, \epos\ and \sibyll). 
  As a benchmark with a standard collider physics model we also show
  \pythia\ (and \phojet) predictions with various parameter settings. 
  While reasonable overall agreement is found for some of the MC, 
  none of them reproduces consistently the $\sqrts$ evolution of all the
  observables. We discuss implications of the new LHC data for the
  description of cosmic-ray interactions at the highest energies. 
\end{abstract}





\end{frontmatter}

%
%

\section{Introduction}

In astroparticle physics, the identification and understanding of the
sources of high-energy cosmic rays is one of the most important open
problems. There is an increasing observational support for the
hypothesis of particles being accelerated at shock fronts in supernova
remnants. These cosmic rays are expected to populate the observed
cosmic-ray spectrum up to $\unit[10^{15} - 10^{17}]{eV}$~\cite{Hillas:2005cs}. The sources
of cosmic rays beyond that energy are not known and, due to the extreme
requirements needed to reach such energies in acceleration processes, subject to 
various speculations including extensions of the Standard Model of particle
physics~\cite{Bhattacharjee:1998qc,Torres:2004hk,Kachelriess:2004ax}.

\begin{figure}[htb]
\centering
\includegraphics[width=0.75\textwidth]{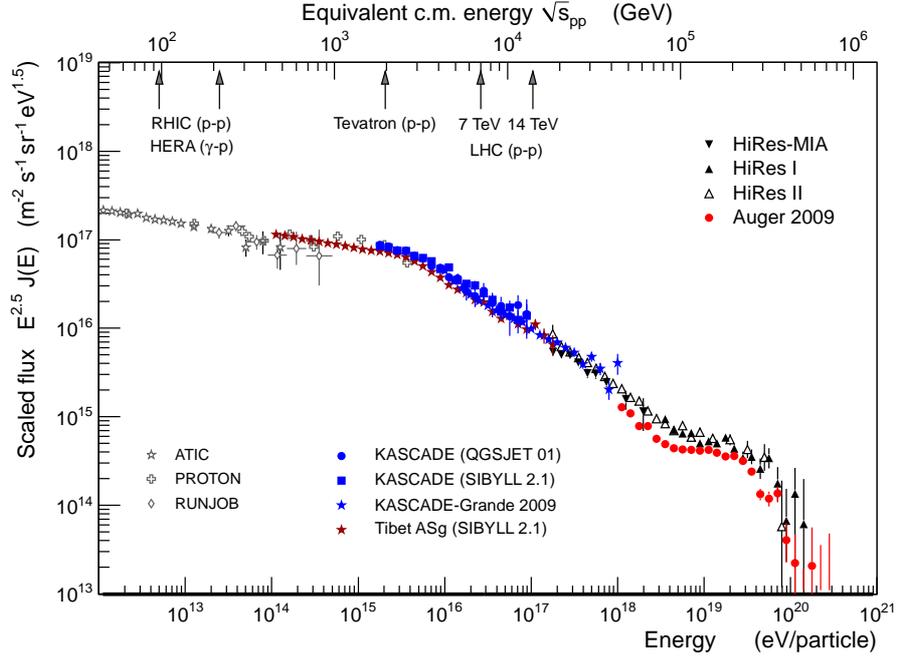}
\caption{Data on the all-particle flux of cosmic rays. The flux has
  been scaled by $E^{2.5}$ to make the features clearly visible. The
  axis at the top indicates the equivalent c.m.~energy if the cosmic
  ray particles were protons. The nominal collider c.m.~energy for
  different accelerators is also shown. References to the data can be
  found in \protect\cite{Bluemer:2009zf}.  }
\label{fig:flux}
\end{figure}

A compilation of recent flux measurements of cosmic rays is shown in
Fig.~\ref{fig:flux}. The power law of the energy spectrum of cosmic
rays changes at about $\unit[10^{15.5}]{eV}$. This feature is known as
the {\em knee}, the origin of which is still under debate. 
Theoretical explanations have been put forward based on 
a change of slope 
in the source spectra, effects of leakage from the Galaxy, the
assumption of changes in hadronic interactions or the production of
exotic new particles~\cite{Hoerandel:2004gv}. The {\em ankle} in the
energy spectrum at $\sim \unit[10^{18.5}]{eV}$ is often assumed to be
the imprint of the change from galactic to extragalactic sources
\cite{Wibig:2004ye,Hillas:2005cs} or, alternatively, a signature of
the propagation of extragalactic cosmic rays through the microwave
background radiation~\cite{Berezinsky:1988wi,Berezinsky:2002nc,Aloisio:2007rc}. 
Finally the suppression of the flux at ultra-high energies could be due to the
Greisen-Zatsepin-Kuzmin (GZK) energy loss effect~\cite{Greisen:1966jv,Zatsepin66e} 
or to the fact that the sources have reached their maximum energy, or to a combination
of both. Indeed, the most powerful astrophysical objects
are expected to be able to accelerate particles up to a maximum energy just around 
$\unit[10^{20}]{eV}$~\cite{Hillas:1985is}. Models based galactic sources, 
which thus elude the GZK suppression, have also been proposed based on UHECR
emission from neutron stars~\cite{Blasi:2000xm,Arons:2002yj} or previous 
Gamma Ray Bursts \cite{Biermann:2004hi,Calvez:2010uh}. In such cases 
the suppression of the spectrum would be related to the maximum energy 
that the sources can accelerate particles to.\\

Knowing the elemental composition of cosmic ray particles arriving at
Earth is of crucial importance to understand the production and
propagation of cosmic rays. Unfortunately, cosmic rays can be measured
only indirectly above an energy of $\unit[10^{14}]{eV}$ through the
cascades of secondary particles, called extensive air-showers (EAS), that 
they produce in the atmosphere (for a recent review, see~\cite{Bluemer:2009zf}). 
Only by simulating the generation of EAS and 
comparing the predictions with measurements one can draw conclusions
on the primary mass composition of the arriving particles~\cite{Knapp:2002vs}. 
With the operation of modern large-scale
experiments the reliability of air-shower simulations has become the
source of the largest systematic uncertainty in the interpretation of
cosmic-ray data~\cite{Antoni:2001pw,Antoni:2005wq,Amenomori:2005nx,%
 AbuZayyad:1999xa,Abraham:2010yv,Abbasi:2009nf}. While the
electroweak interaction processes are reasonably well
understood, modeling of hadronic multiparticle production is subject
to large theoretical uncertainties that are difficult to
estimate~\cite{Knapp96a,Zha:2003bt,Ulrich:2010rg}.\\

In this context it is not surprising that in some speculative scenarios the knee in
the cosmic-ray spectrum has been related to a change in the
characteristics of hadronic interactions. In an air-shower only a part
of the energy of the primary particle is transferred to
electromagnetic particles and low-energy muons that can be detected
easily. Assuming that, for example, a rapidly increasing fraction of
very high energy muons is produced in interactions just above the knee
energy, the non-detection of the energy of these muons would result in
a systematically incorrect reconstruction of the primary energy and
lead to a knee in the observed spectrum even if there is not such a 
break in the primary CR flux~ (see, for example,
\cite{Petkov:2005gv,Petrukhin:2006kp,Barcelo:2009uy,Dixit:2009mt}).
 These models could not
be constrained by Tevatron data as the $\unit[2\times 10^{15}]{eV}$ 
equivalent (fixed-target) energy of the Tevatron collider is below the knee.\\

The Large Hadron Collider (LHC) at the CERN laboratory allows us to access for the first time the energy region
above the knee in the laboratory. Therefore an analysis of inclusive particle 
data taken at the LHC is particularly interesting for constraining
existing hadronic interaction models and for testing possible new mechanisms
of hadron production~\cite{Alessandro:2011wt}.
Data from LHC experiments published so far have mostly been taken with
detectors covering the central phase space region in pseudorapidity ($|\eta|\lesssim$~2.5). 
This region is most easily accessible in collider
experiments and is also the region of the highest rapidity-density of
produced particles. On the other hand, since the number of particles
in an air-shower is roughly proportional to the energy of the primary
particle, the most energetic outgoing particles of an interaction, 
emitted in the very forward region of a collider experiment 
-- such as in diffractive interactions -- are
the most important ones for understanding air-showers. 
While waiting for the results from forward detectors at the LHC~\cite{Engel:2004is,Sako:2007zz,dEnterria:2008jk},
important constraints can be derived
from the already measured central particle production.\\

In this paper we compare the predictions of several representative
hadronic interaction models with single-particle inclusive observables
measured at midrapidity at the LHC. We focus on the following observables:
(i) the pseudorapidity density of charged particles at midrapidity $\dNdeta$, 
(ii) the average transverse momentum of hadrons $\meanpt$; and (iii) the
event-by-event distribution of the charged multiplicity,
$P(N_{ch})$. Implications of the data-theory comparisons are discussed and
further observables of direct relevance to EAS simulations, 
measurable with central and forward detectors at the LHC, are pointed out.\\

The paper is organized as follows. In Section~\ref{sec:th} we recall
the main theoretical ingredients of the hadronic interaction models
used in our study and discuss their differences. In
Section~\ref{sec:data}, we summarize the recently collected inclusive 
charged hadron data at the LHC from the ALICE~\cite{Aamodt:2010ft,Aamodt:2010pp} 
and CMS~\cite{Khachatryan:2010xs,Khachatryan:2010us} experiments as well 
as results from previous 
colliders (ISR~\cite{Rossi:1974if}, UA1~\cite{Albajar:1989an},
UA5~\cite{Alner:1986xu} and CDF~\cite{Abe:1989td,Abe:1988yu}).
[Unfortunately, we could not include in this study the latest results from 
the ATLAS experiment~\cite{Aad:2010ir}, which appeared public only when 
this analysis was being finished.]
We discuss the different event trigger selection criteria applied
in each experiment as those affect somewhat the data-model
comparisons. In Section~\ref{sec:data_th} we compare the measurements
to the various event generators, and discuss the results in the context
of high energy cosmic-ray interactions with the atmosphere in
Section~\ref{sec:discuss}. We summarize our main findings in
Section~\ref{sec:summ}.

%
%

\section{High-energy hadronic interaction models}
\label{sec:th}

Calculating predictions for the bulk of produced particles in hadronic
interactions is not possible yet within first-principles Quantum-Chromodynamics (QCD).
One has to resort to phenomenological models that combine
fundamental principles of quantum field theory -- such as unitarity, analyticity and
crossing -- together with perturbative QCD (pQCD) predictions including phenomenological 
fits (e.g.~accounting for the parton-to-hadron fragmentation) to experimental hadron spectra.\\ 

General-purpose hadronic interaction models used in high-energy physics
(HEP), such as \pythia~\cite{Sjostrand:2006za}, \herwig~\cite{Corcella:2000bw} 
and \textsc{sherpa}~\cite{Gleisberg:2008ta}, are developed to learn and interpret 
the data measured in accelerator experiments with an emphasis on hard-scattering
measurements (signals and backgrounds) rather than on the bulk of hadron production
at lower transverse momenta. The predictions of models can be adjusted by
a large number of parameters that encode fundamental physics quantities
on the one hand, and phenomenological assumptions or simplifications on
the other. These models are typically optimized for processes
that can be calculated in perturbation theory and other assumptions
needed for generating complete final states for hadronic interactions
are often kept at a more simple level. In addition, these models are
mainly designed for proton, pion or photon interactions and not suited
for nuclear interactions relevant for CRs collisions with air nuclei
in the upper atmosphere.\\ 

In contrast, interaction models commonly used in cosmic-ray physics such as
\qgsjet 01~\cite{Kalmykov:1997te,Kalmykov:1993qe}, 
\qgsjet II~\cite{Ostapchenko:2005nj,Ostapchenko:2004ss,Ostapchenko:2007qb} and
\sibyll~\cite{Engel:1992vf,Fletcher:1994bd,Ahn:2009wx},
are supposed to predict hadronic interactions as realistically as
possible with the emphasis on reproducing existing accelerator
measurements and providing a reasonable extrapolation to higher energy
and to phase-space regions where no data are available. 
On the one hand, there is typically only one ``optimal''
parameter set for each of the models and simplifications are made in
the implementation of known QCD predictions, in particular, if they are
not directly relevant to the production of high energy
secondaries. On the other hand, more sophisticated models of soft particle production
are implemented and great care is given to the relation of the total,
elastic and inelastic cross sections to particle production.\\

In between these two generic categories there are models such as
\phojet~\cite{Engel:1994vs,Engel:1995yd,Engel:1995sb}/\dpmjet~\cite{Roesler01a,Bopp:2005cr} 
and \epos~\cite{Werner:2005jf}, which are designed to be more universal and approach the
sophistication of HEP models regarding some aspects of hard processes. At the same
time the parameters of these models are tuned to describe a large set
of accelerator data at various energies and are not supposed to be
changed for tuning for different experiments.\\

In the following we will consider the models \dpmjet, \qgsjet 01 and
II, \sibyll\ 2.1, and \epos~v1.99 which are often used in cosmic-ray
simulations~\cite{Knapp:2002vs,Bluemer:2009zf} and the \pythia\ Monte
Carlo versions 6.4 and 8 as a benchmark HEP model. The \pythia\
predictions are shown mainly for reference as this model is used in
all publications on minimum-bias measurements at the LHC. Since the
proton-proton interaction generator in \dpmjet III is identical to
\phojet\ we will use the latter in our comparisons.\\

Providing a description of the physics implemented in these models
would be beyond the scope of this work. Here we will rather recall
only the basic concepts behind the modeling particle production. 

\subsection{Modeling of central multiparticle production} 

The inclusive production of particles in high-energy hadronic
collisions receives contributions from ``soft'' and ``hard''
interactions between the partonic constituents\footnote{Strictly speaking 
one cannot refer to individual partons in case of soft interactions because 
of confinement. Still it is a practical approach taken in all event generators 
to extend some of the concepts of hard interactions to the soft domain.} 
of the colliding hadrons. Soft (resp. hard) processes involve mainly $t$-channel 
partons of virtualities $q^2$ typically below (resp. above) a scale
$Q_0^2$ of a few~GeV$^2$. Soft scatterings give rise to production of
hadrons with low transverse momenta $\pT$ and dominate hadronic
collisions at low energies ($\sqrts\lesssim$~20~GeV).\\

Although soft processes have a virtuality scale not far from
$\Lambdaqcd \approx$~0.2~GeV and thus cannot be treated within pQCD,
predictions based on basic quantum field-theory principles such as
unitarity and analyticity of scattering amplitudes, as implemented in
the Gribov's Reggeon Field Theory (RFT)~\cite{Gribov:1968fc}, give a
decent account of their cross sections in terms of the exchange of
virtual quasi-particle states (Pomerons and Reggeons).
At increasing energies, the Pomeron contribution, identified perturbatively with a multi-gluon exchange, 
dominates over those from secondary Reggeons (virtual mesons). In particular, the {\it soft} Pomeron exchange, 
of colourless nature, is responsible for particle production in peripheral hadronic collisions 
and, in particular, for diffractive dissociation which accounts for a noticeable fraction of the 
total inelastic cross section all the way up to asymptotic energies.\\

(Semi)hard parton-parton scatterings dominate the inelastic hadron production cross-sections 
for c.m.~energies above a few hundreds of GeV. Hard processes with large $|q^2|\gg \Lambdaqcd^2$ 
can be treated within pQCD in a collinear-factorized approach where one sums up Feynman 
diagrams of the underlying parton-parton (quarks and gluons) scatterings, which give the 
leading logarithmic contribution, with each power of the strong coupling constant $\alpha_s(q^2)$ 
being multiplied by a collinear logarithm $\log q^2$. 
The scattered quarks and gluons produce then collimated bunches of final-state hadrons
(jets) in a branching process dominated by perturbative parton splittings 
described by the Dokshitzer-Gribov-Lipatov-Altarelli-Parisi (DGLAP) equations~\cite{Gribov:1972ri,Altarelli77,Dokshitzer77}, 
and by non-perturbative hadronization when the parton virtuality is below $\cO{1~GeV}$.
The RFT approach can also be generalized to include hard processes via ``cut (hard) Pomerons'' diagrams.
The physical picture is that of a colour flux tube, which is in the simplest case treated as a classical 
string that subsequently fragments into hadrons.\\

Historically, Monte Carlo (MC) event generators of high-energy hadronic collisions have evolved 
either starting up from the RFT approach, 
exemplified e.g.~in the original Dual Parton Model (DPM)~\cite{Capella:1992yb}, 
extended with a leading-logarithmic pQCD description for high-$\pT$ production 
(based on cut-Pomerons) -- such as in the \phojet~\cite{Engel:1994vs,Engel:1995yd}, \qgsjet 01 and
II~\cite{Kalmykov:1997te,Ostapchenko:2005nj,Ostapchenko:2004ss,Ostapchenko:2007qb},
\sibyll~\cite{Engel:1992vf,Fletcher:1994bd,Ahn:2009wx},
\nexus~\cite{Drescher:2000ha,Pierog:2002gj},
\epos~\cite{Werner:2005jf} and \dpmjet~\cite{Roesler01a,Bopp:2005cr} cases -- or they started 
from a purely collinear-factorized framework -- such as in e.g.\ general-purpose MCs like 
\pythia~\cite{Sjostrand:2006za} -- 
complemented with an add-on model for truly soft~\cite{Sjostrand:1987su} 
and diffractive~\cite{Schuler:1993wr} scatterings. 
Thus, on the one hand, the RFT approaches try to extend a consistent framework based on Pomeron 
degrees of freedom to the hard regime. On the other, the collider MCs contain a description 
based on partonic degrees of freedom (with scattering cross sections dumped in the infrared, 
below a ``tunable" semihard scale) with soft and diffractive scatterings 
incorporated in a more or less {\it ad hoc} way.
In both approaches the final non-perturbative transition of partons 
to hadrons is modeled based on the ideas of the Lund string fragmentation model~\cite{Andersson:1983ia}.
At increasingly higher $\sqrts$, in both frameworks one has to account for multiple scattering 
processes between the colliding hadrons, 
namely one has to include multi-Pomeron exchanges and/or 
multiple hard scattering processes.\\ 

In the RFT framework, the {\it single} Pomeron ($\Pom$) exchange amplitude is characterized by 
a power-like energy dependence, $f^{\mathbb P}(s,t)\propto s^{\alpha _{\mathbb P}(0)}$,
with the Pomeron intercept $\alpha _{\mathbb P}(0)\sim 1.1$ leading to a corresponding 
energy rise of the total cross section $\sigma _{\rm tot}=\frac{1}{2s}\,{\rm Im} f^{\mathbb P}(s,0)$,
which asymptotically violates the so-called Froissart bound ($\sigma_{tot}< c\,\log^2 s$)~\cite{Froissart:1961ux}.
Accounting for eikonal multi-Pomeron exchanges, the cross sections are unitarized, i.e.\
$\sigma _{\rm tot,inel} \propto \log^2 s$, although due to the Abramovskii-Gribov-Kancheli (AGK) 
cancellations~\cite{Abramovsky:1973fm} such multi-Pomeron configurations give zero contribution to 
inclusive particle spectra. Thus, the total soft charged particle density produced at midrapidity 
follows the energy-dependence defined by a single Pomeron exchange contribution:
\begin{equation}
\left.\frac{dN_{{\rm ch}}(s,\eta)}{d\eta}\right|_{\eta=0} \propto 
\frac{{\rm Im} f^{\mathbb P}(s,0)}{s\:\sigma_{pp}^{{\rm inel}}(s)}\;
\sim\frac{s^{\Delta_{\mathbb{P}}}}{\log^{2}s}\,,\;\;\;\;\mbox{ with $\Delta_{\Pom}\equiv \alpha_{\Pom}(0)-1 \sim 0.1$.}
\label{eq:dNsoft}
\end{equation}

In pure DGLAP-based models, the central pseudo-rapidity particle density is proportional to the 
inclusive jet cross section which is given by the convolution of parton distributions 
functions (PDFs) and parton-parton scattering cross sections:
\begin{eqnarray}
\sigma_{pp}^{{\rm jet}}(s,Q_{0}^{2}) & = & \int dx_{1}\,dx_{2}\,\int\! d\pT^{2}\;
\sum_{i,j=q,\bar{q},g} f_{i/p}(x_{1},\pT^{2})\;f_{j/p}(x_{2},\pT^{2}) 
\times \frac{d\sigma_{ij}(x_{1}x_{2}s,\pT^{2})}{d\pT^{2}} 
 \;\;\Theta(s-4\pT^{2})\,.
\label{eq:sigma_jet}
\end{eqnarray}
The hard cross section is divergent in the limit $\pT\to 0$ and one needs to introduce a $\pT$-cutoff 
$Q_{0}$ to indicate the regime of validity of the perturbative approximation. At increasingly larger
c.m.~energies, one needs to account for multi-parton scatterings and saturation effects. On the one hand,
the cross section predicted by the regularized processes exceeds the total inelastic cross section, 
indicating that several (or multiple) hard scatterings occur per collision. On the other,
for decreasing but still perturbative $\pT$ values, the integrals receive major contributions from 
the region of low parton fractional momenta ($x=p_{\mbox{\tiny{\it parton}}}/p_{\mbox{\tiny{\it hadron}}}$), 
where the dominant gluon distribution rises
roughly as $f_{g/p}(x,\pT^{2})\sim x^{-\Delta_{{\rm hard}}}$
with $\Delta_{{\rm hard}} \simeq 0.3$. 
After integrating above the $\pT$-cutoff $Q_{0}$, 
one obtains an energy-dependence of the corresponding hard central charged hadron densities of the type
\begin{eqnarray}
\left.\frac{dN_{{\rm ch}}(s,\eta)}{d\eta}\right|_{\eta=0} & \sim & \frac{\sigma_{pp}^{{\rm jet}}(s,Q_{0}^{2})}{\sigma_{pp}^{{\rm inel}}(s)}
\sim\frac{s^{\Delta_{{\rm hard}}}}{Q_{0}^{2}\:\log^{2}s}\,,
\;\;\;\;\mbox{ with $\Delta_{\rm hard} \approx 0.3$.}
\label{eq:dNhard}
\end{eqnarray}
Clearly, the fast growth of the gluon densities at low $x$ results in the hard part of
the particle density ($\propto s^{\Delta_{{\rm hard}}}$, $\Delta_{{\rm hard}}\sim 0.3$) 
to rise with energy much faster than for soft processes ($\propto s^{\Delta_\Pom}$, $\Delta_\Pom\sim 0.1$).
However, at sufficiently small $x$, the number of gluons is so large that new 
parton multiscattering phenomena have to be accounted for. First, non-linear ($gg$ fusion) 
effects become important in the PDFs themselves, saturating their growth as $x\to$~0~\cite{Gribov:1984tu}.
The strength of these effects is controlled by the ``saturation scale'' $Q_{\rm sat}^2$
at which parton branching and fusion processes start to compensate each other. 
Second, the probability to have simultaneous scatterings of the constituents of the
colliding hadrons also increases leading to multiple parton interactions (MPI) in a single collision.
In many MC generators one effectively mimics 
saturation effects by introducing 
some energy dependence to the infrared $\pT$-cutoff: $Q_{0}^2=Q_{0}^2(s).$
In particular, choosing $Q_{0}^{2}(s)\sim s^\varepsilon$ as done 
e.g.~in \pythia, the power-law changes as
\begin{equation}
\left.\frac{dN_{{\rm ch}}(s,\eta)}{d\eta}\right|_{\eta=0}\sim
 s^{\Delta_{{\rm hard}}-\varepsilon}\;,
\label{eq:12}
\end{equation}
which reduces the relative role of the hard contributions compared to the ``linear'' 
predictions for fixed $Q_0$ cutoff, Eq.\ (\ref{eq:dNhard}). [We note that {\it stricto sensu}
in \pythia\ there is extra particle production from ``purely" soft scatterings~\cite{Sjostrand:1987su} 
which is not accounted for by this simple expression.]\\

Not only the particle multiplicities but also their transverse momenta are sensitive to the
underlying parton dynamics. In soft processes, the produced hadrons have typical
transverse momenta in the sub-GeV range ($\pT^{\rm soft}\sim 0.3 - 0.4$ GeV),
without a pronounced energy-dependence for the average $\pT$. For the hard contribution, 
due to the fast drop ($\propto 1/\pT^4$) of the differential jet production cross-section,
the peak of the perturbative production sits at gluons whose transverse momentum is
close to the cutoff, $\pT \sim Q_0$, producing jets of a few GeV which are often called ``minijets'' since they are 
perturbatively calculable but often not experimentally observable over the soft hadronic background. 
Due to the faster energy rise of the 
hard contribution compared to the soft one -- cf.\ Eqs.\ (\ref{eq:dNsoft}) and (\ref{eq:dNhard}) --
average hadron transverse momenta rise with $\sqrts$ from $\pT\sim \pT^{\rm soft}$ to $\pT\sim Q_0$. 
As a matter of fact, the energy-dependence of the average $\pT$ is expected to be even steeper when parton
saturation effects come into play -- due to the suppression of semi-hard parton production
in the ``dense'' saturated region, $q^2\simeq\pT^2 < Q_{\rm sat}^2(x)$ and 
due to the rise of the 
saturation scale $Q_{\rm sat}^2\propto\log(1/x)\propto\log(\sqrts)$ itself. 
The effective value of $\langle Q_{\rm sat}^2\rangle$ controls thus the mean transverse momentum 
$\meanpt$ of a significant part of the finally produced hadrons after parton fragmentation.
It is noteworthy that the MC procedure of mimicking the saturation and MPI effects via an 
energy-dependent $\pT$-cutoff changes somehow this picture. Having the soft part of the interaction
unchanged and reducing the minijet contribution -- cf.\ Eqs.\ (\ref{eq:12}) and
(\ref{eq:dNhard}) -- the $\pT$-rise from $\pT^{\rm soft}$ to $Q_0(s)$ proceeds
slower than in the linear case, up to very high energies.\\ 

All in all, it is clear that both the total particle pseudorapidity density $\dNdeta$ 
and the average $\pT$ of produced hadrons are sensitive to the soft and semi-hard dynamics, 
to the non-linear parton (saturation) effects, and to their implementations in the models of 
high-energy hadronic scattering.\\

\subsection{The \pythia\ event generator}

The basic ingredients of the \pythia\ event generator are leading-order pQCD 2 $\to$ 2 matrix elements, 
augmented by initial- and final-state parton showers (ISR and FSR, respectively), folded with 
parton distribution functions (interfaced here via the \lhapdf\ v5.8.2 package~\cite{Whalley:2005nh}) 
on the initial-state and the Lund string model~\cite{Andersson:1983ia} to describe the final 
parton-to-hadron fragmentation. The infrared $1/\pT^4$ divergence of the hard (multi)parton cross sections 
is regularized by a cutoff $Q_0$, such that $1/\pT^4 \to 1/(\pT^2+Q_0^2)^2$. The infrared cutoff
depends on the hadron-hadron c.m.~energy: 
$Q_0^2(s)=Q_0^2(s_0)\cdot(s/s_0)^\epsilon$, where $Q_0(s_0)$ is a reference value 
at a given c.m.\ energy $\sqrt{s_0}$, e.g.~$Q_0\approx$~2~GeV at $\sqrt{s_0}$~=~1.8~TeV 
is preferred by the Tevatron data~\cite{Affolder:2001xt,Acosta:2004wqa}.
Other non-perturbative ingredients of \pythia\ include a Regge-based modeling of diffractive 
processes~\cite{Schuler:1993wr}, plus a model for the underlying-event (UE) issuing from 
multi-parton interactions (MPI), soft scatterings, and beam-remnants~\cite{Sjostrand:1987su}.
Multiple parton collisions are treated ``perturbatively'' -- i.e.~based on the eikonalization of {\it hard} scattering 
contributions -- but require a non-perturbative ansatz for the impact-parameter 
profile of the colliding hadrons.\\

In this work we use the \pythia\ event generator in two flavours: the Fortran version 6.422~\cite{Sjostrand:2006za}
(Nov. 2009), as well as the C++ version \pythia\ 8.130~\cite{Sjostrand:2007gs}. Both codes
include the newest description of multiple parton interactions based on the concept of 
``interleaved evolution'' given by $\pT$-ordered 
showers~\cite{Sjostrand:2004pf,Sjostrand:2004ef} which accounts well for 
correlations between the hard-scattering and the underlying-event components seen in the Tevatron data.
Compared to \pythia~6.4, the MPI, ISR and FSR in \pythia\ 8 are all interleaved, 
and the treatment of diffraction has improved:
a diffractive system is viewed as a Pomeron-proton collision, 
including hard scatterings subject to all the same ISR/FSR 
and MPI dynamics as for a ``normal'' parton-parton process.\\

We consider three different ``tunes'' of the non-perturbative ingredients 
(IS and FS showering, UE, beam-remnants, FS colour-reconnection, and hadronization)
of the two versions of \pythia.
For \pythia\ 6.4 we selected (via the \ttt{PYTUNES} switch) the
so-called Perugia-0 (320)~\cite{Skands:2009zm,Skands:2010ak} 
and Atlas-CSC~(306)~\cite{Moraes:2009zz,Bartalini:2010su} tunes. 
Comparisons of LHC data to other existing \pythia\ 6 tunes can be found in the literature
(see e.g.~\cite{Field:2010bc}).
For \pythia\ 8 we use the default ``tune 1''. 
The chosen settings describe one or more of the minimum-bias (MB), underlying-event (UE), and/or Drell-Yan (DY) 
data sets at the Tevatron~\cite{Skands:2007zz,Skands:2009zm,Skands:2010ak}, and are reasonably complementary on a number 
of important points, as illustrated in Table~\ref{tab:pythiaTunes}.\\

\begin{table}[htbp]
\begin{center}
{\footnotesize
\begin{tabular}{lcccccccc}\hline
Version & Tuning  & Diffraction & $Q_0$ cutoff at & $Q_0$ scaling & PDF & Proton  & FS colour & Exp. constraints \\
        & (\ttt{PYTUNES}) &  & $\sqrt{s_0}$~=~1.8 TeV & power $\epsilon$ & & profile & reconnection & (Tevatron)\\\hline
6.422 & Perugia~0~(320) & simple & 2.0 GeV & 0.13 & CTEQ5L & ExpOfPower & weak &   UE, MB, DY \\
6.422 & Atlas-CSC~(306) & simple & 1.9 GeV & 0.11 & CTEQ6L & double-Gauss & weak & MB, DY \\
8.130 & Tune 1 & improved  & 2.15 GeV & 0.08 & CTEQ5L & double-Gauss & weak & MB \\\hline
\end{tabular}
}
\caption{Comparison of the various ingredients controlling the non-perturbative and semi-hard dynamics in
 the three \pythia\ models used in this work. See text for details.} 
\label{tab:pythiaTunes}
\end{center}
\end{table}

Since the three tunes share the same MPI dynamics, the single most important parameter to extrapolate 
the hadron multiplicity from Tevatron to LHC energies is the exponent $\epsilon$ which controls 
the dependence of the MPI infrared cutoff scale on the collision energy, see Eq.~(\ref{eq:12}). 
Given that a single value of $Q_0$ is used to regularize both the hard scattering 
and the MPI in the event, 
and that the generation of additional parton-parton interactions in the underlying
event is suppressed below $Q_0$, a {\it higher} scaling power of the infrared cut-off 
implies a {\it slower} increase of the overall hadronic activity.
The default (``untuned'') \pythia\ 6.4 energy scaling\footnote{Parameter \ttt{PARP(90)} = $2\cdot\varepsilon$ = 0.16.},
$\epsilon = 0.08$, was chosen to follow the scaling of the total cross section ($\Delta_{\Pom}\sim 0.08$).
However, most of the subsequent tunes needed to significantly increase the scaling power value to $\epsilon \approx 0.13$,
in order to find a better agreement with the $\sqrts$-dependence of the UE and MB measured from Sp$\bar{\mbox p}$S 
and Tevatron~\cite{Skands:2009zm,Skands:2010ak}. One of the first conclusions from the MB LHC data, 
as we will see next, is that this exponent needs to be lowered again.\\

For the initial-state Atlas-CSC uses the CTEQ6L parton densities~\cite{Pumplin:2002vw} plus a
double-Gaussian transverse distribution for the partons in the proton, whereas Perugia 
uses the default CTEQ5L PDFs~\cite{Lai:1999wy} and a smoother proton form-factor (exponential-of-power, 
$\exp(-r^{n})$, with exponent $n$~=~1.7) which decreases the multiplicity fluctuations. 
The Tune-1 of \pythia\ 8 is closer to Atlas-CSC in terms of the energy-scaling of the pQCD cut-off 
(it features the smallest exponent, $\varepsilon$~=~0.08, of the three models) and of the proton profile
but uses the default CTEQ5L PDFs. 
For the final-state, the three tunes have weak final-state colour reconnections, which {\it increase} 
the final particle multiplicities, compared to models with stronger colour correlations.

\subsection{Models based on Reggeon Field Theory (RFT)}

The RFT-based MCs considered here start off from a construction of the hadron-hadron elastic 
scattering amplitude to determine the total and elastic cross sections. 
Experimental data allow one to fix the basic model parameters, such as the intercept and 
slope of the Pomeron Regge trajectory, the Pomeron-hadron couplings, etc. 
The inelastic events are then understood in terms of cut Pomerons which correspond 
to 
colour flux tubes (treated as strings) extended between constituent partons of the colliding
hadrons. 
Final hadrons are produced in the fragmentation of the strings.
The different models used in this work (Table~\ref{tab:MCs}) differ in various approximations
for the collision configurations (e.g.\ the distributions for the number of cut 
Pomerons and for the energy-momentum partition among them), the treatment of diffractive 
and perturbative contributions as well as of high parton density effects, and the 
details of particle production from string fragmentation.\\


\begin{table}[htbp]
\begin{center}
\begin{tabular}{lcccc}\hline
Model (version)             & Diffraction         & $Q_0$ cutoff & Saturation       & Other \\
                            &                     &   (GeV)      & effects         &       \\ \hline
\phojet\ 1.12~\cite{Engel:1994vs,Engel:1995yd} & 2-channel eikonal   & 2.5  & via running $Q_0(s)$ & FS+hadronization via \pythia\ 6.115 \\
\qgsjet 01~\cite{Kalmykov:1997te} & quasi-eikonal       & 2.0   & none (flat low-$x$ PDFs) &   --\\
\sibyll\ 2.1~\cite{Engel:1992vf,Fletcher:1994bd,Ahn:2009wx} & 2-channel eikonal   & 1.0   & via running $Q_0(s)$    & --\\
\qgsjet II~\cite{Ostapchenko:2005nj,Ostapchenko:2004ss,Ostapchenko:2007qb} & cut enhanced graphs & 1.6   & enhanced $\Pom$-graphs  & --\\
\epos\ v1.99~\cite{Werner:2005jf}    & diffractive Pomeron & 2.0   & parametrized           & final-state collective flow \\\hline
\end{tabular}
\caption{Comparison of the main ingredients controlling the non-perturbative and semi-hard dynamics
present in the RFT-based event generators used in this work.}
\label{tab:MCs}
\end{center}
\end{table}

Whereas the Regge-Gribov approximation is applied to hadrons as interacting objects in the case 
of \sibyll, \qgsjet\ and \phojet, it is extended to include partonic constituents in \epos. 
In the latter case, this allows the implementation of energy sharing between the different constituents of a hadron 
at amplitude level and suppresses final states with very large particle multiplicity.\\

Except in \epos, the basic implementation of diffraction dissociation follows the Good-Walker
formalism~\cite{Good:1960ba} where the colliding hadrons are represented by superpositions of 
elastic scattering eigenstates which undergo different absorption during the collision. 
In all these models one uses two eigenstates per hadron (2-channel eikonal approach);
in \qgsjet 01 the second eigenstate is ``passive'' (has zero coupling to Pomerons) which leads to
the so-called quasi-eikonal approach~\cite{Kaidalov:1979jz}. While in \sibyll\ v2.1 this mechanism is
associated with high-mass diffraction, in the others it is restricted to the production of low-mass diffractive states.
High-mass diffraction is treated empirically in \qgsjet 01, 
accounting for single diffraction only. It is based on diffractive cuts of triple-Pomeron and Pomeron-loop 
graphs in \phojet, 
and it is based on all-order resummation of cut enhanced $\Pom$-graphs in \qgsjet II.
In \epos\ v1.99, a special kind of Pomeron is used to 
define a diffractive event. Depending on each event configuration it can be 
non-diffractive, low mass diffraction without central particle production, 
or high mass diffraction. In all the cases, each initial hadron can be in an excited state.\\

All the models account for multiple soft and (semi)hard production processes via multi-Pomeron
interactions. 
The effects of high parton density at small $x$ and the treatment of the hadronic remnants 
are implemented differently in the different generators.
\sibyll\ 
uses an energy-dependent transverse 
momentum cutoff for minijet production, $Q_0(s)\sim Q_0 + \exp{(\log s)}$, 
based on the geometric criterion that there cannot be more gluons in a hadron than would 
fit in a given transverse area~\cite{Bopp:1994cg}. 
This cutoff is independent of the impact parameter of the collision. 
\phojet\ uses also a $\sqrts$-dependent cutoff, of the form $Q_0(s) \sim Q_0 + C \cdot \log(\sqrts)$.
\qgsjet 01 is based on old parton densities of the pre-HERA era and
the (artificial) flatness of low-$x$ PDFs, $x\,g(x,Q_0^2)\propto \rm{const}$,
allows one to partly avoid the problem of high parton densities. This issue is addressed 
more realistically in the successor model, \qgsjet II, which is based on a re-summation
of enhanced diagrams corresponding to Pomeron-Pomeron
interactions~\cite{Ostapchenko:2005nj,Ostapchenko:2006vr,Ostapchenko:2008},
which lead to impact-parameter and density-dependent
parton saturation for soft processes~\cite{Ostapchenko:2005yj}.
In \epos, non-linear effects are introduced phenomenologically via the
elastic and inelastic fusion of parton-ladders. 
The elastic process provides screening, therefore a reduction of total
and inelastic cross sections. The inelastic process affects particle
production, in particular transverse momentum spectra.
Finally, in contrast to other MCs, 
the \epos\ version 1.99 used in this paper includes also a parametrized\footnote{The latest EPOS
developments (\epos 2)~\cite{Pierog:2010dt,Werner:2010ss} include a full hydrodynamical evolution.} 
collective expansion in the final state for all colliding systems including \pp\ at LHC energies.

%
%

\section{Experimental data and event selection corrections}
\label{sec:data}

Proton-(anti)proton interactions at colliders can be roughly divided into three 
categories according to the overall event topology: elastic, diffractive and 
non-diffractive processes. Elastic events are usually missed by the experiments 
unless specific detectors (Roman Pots) are installed inside the tunnel~\cite{Anelli:2008zza}. 
Inelastic interactions are often collected with a generic minimum bias (MB) trigger 
that accepts a large fraction of the particle production cross section by requiring a minimum 
activity in one or various detectors\footnote{True ``zero bias'' triggers, which only require 
the passing by of the two beams by the interaction point, are more sensitive to instrumental 
backgrounds.}. Often one requires a two-arm trigger with particles in opposite 
hemispheres to eliminate backgrounds mostly from beam-gas collisions and from cosmic-rays.
Such a trigger selection, however, reduces significantly the detection rate of (single) diffractive
collisions characterized by the survival of one of the colliding (anti)protons 
and particle production in just one hemisphere. The associated measurements are 
often dubbed ``non single-diffractive'' (NSD), although the corresponding cross
sections are obtained with some level of model-dependence as different
event-generators have different descriptions of diffractive production. 
The UA5 experiment triggered only on one arm to measure the ``single-diffractive'' (SD)
events and by combining the SD and the NSD results published MB results very close to 
a true inelastic trigger (INEL) (using MC corrections in both cases). LHC experiments define their
inelastic trigger in different ways. \\

The list of minimum-bias (non-single diffractive and inelastic) event-selection triggers that have been employed 
by the different experiments presented in this work are listed in Table~\ref{tab:triggers}.
Since in most cases, and in particular at the LHC, the experimental MB-trigger corrections have been obtained using \pythia\
and \phojet, we can compare the MC-truth results from these two generators {\it directly} to the data. 
For the RFT models, we will use the predictions obtained both by using a theoretical definition
of an NSD event as well as by applying directly the experimental cuts at the hadron-level to the final states 
produced in the MC simulation. This will allow us to estimate the effect of the trigger definitions 
on the measured hadron distributions and to investigate the model dependence of the experimental 
event-selection corrections.\\

\begin{table}[htbp]
\begin{center}
\begin{tabular}{l c c c}\hline
Experiment \hspace{0.1cm} & \hspace{0.3cm} Collision  \hspace{0.3cm} &
Non-single diffractive (NSD)  & Inelastic (INEL)  \\
 \hspace{0.1cm} & \hspace{0.3cm}  $\sqrts$  \hspace{0.3cm} &  trigger &  trigger \\\hline
\multirow{2}{*}{{\bf ISR}} & \pp & $-$ & Cross section based\\
 & 30 - 65 GeV & &  on luminosity \\
\multirow{2}{*}{{\bf UA1}} & \ppbar\ & 1 or more charged particles at each & $-$ \\ 
 & 0.2 - 0.9 TeV & opposite rapidities ($1.5 < |\eta| <5.5$) & \\
\multirow{2}{*}{{\bf UA5}} & \ppbar\ & 1 or more charged particles at each&  1 or more charged particles \\ 
 & 0.2 - 0.9 TeV & opposite rapidities ($2.0 < |\eta| <5.6$) & within $2.0 < |\eta| <5.6$ \\
\multirow{2}{*}{{\bf CDF}} & \ppbar\ & 1 or more charged particles at each& $-$ \\ 
 & 1.96 TeV & opposite rapidities ($3.2 < |\eta| <5.9$) &  \\
\multirow{2}{*}{{\bf ALICE}} & \pp & 1 or more tracks at each opposite & 1 or more hits at $|\eta| <2$ \\ 
 & 0.9 - 7 TeV & rapidities ($2.8 < \eta < 5.1$, -$1.7 < \eta <-3.7$) &  OR $2.8 < \eta < 5.1$ OR -$1.7 < \eta <-3.7$\\
\multirow{2}{*}{{\bf CMS}} & \pp & \hspace{0.1cm} 1 or more particles ($E_{i}> 3$~GeV) at each \hspace{0.1cm} &
\hspace{0.5cm} 1 or more charged particle(s) \hspace{0.5cm}\\ 
 & 0.9 - 7 TeV & opposite rapidities ($2.9 < |\eta| <5.2$) & in the central region $|\eta| < 2.0$ \\\hline
\end{tabular}
\caption{List of hadron-level cuts implemented in the ``non-single diffractive'' (NSD) and ``inelastic'' (INEL)
triggers used by various proton-(anti)proton collider experiments.} 
\label{tab:triggers}
\end{center}
\end{table}

Primary charged hadrons are defined as all charged hadrons produced in the collision,
including the products of strong and electromagnetic decays, but excluding products of 
weak decays. Feed-down corrections from weakly decaying strange resonances (mainly $\mathrm{K^0_S}$, 
$\Lambda,\overline{\Lambda}$ and $\Sigma^+,\overline{\Sigma}$) have to be accounted for 
in order to obtain the final hadron spectrum. Such corrections, which depend 
on the strange particle composition in the MC, reduce by about $\sim$8\% the total charged yield 
at midrapidity. In all the simulations, one takes this into account by decaying all unstable 
particles for which\footnote{\ttt{MSTJ(22)=2,PARJ(71)=10} in \pythia\ 6.4, and 
\ttt{ParticleDecays:limitTau0 = on, ParticleDecays:tau0Max = 10} in \pythia\ 8.} 
$c\tau<$~10~mm.
The sole contribution from charged leptons to the reconstructed tracks in the 
low-$\pT$ range, comes from the Dalitz $\pi^0$ decay amounting to about 1.5\% of 
the charged yield. ALICE does not correct for this contribution, whereas CMS does. 
We have removed this small contribution from all our model predictions by 
counting only the produced charged hadrons.

%
%

\section{Data versus models}
\label{sec:data_th}

%

\subsection{Particle pseudorapidity densities}
\label{sec:dNdeta}

\begin{figure}[htbp]
\includegraphics[width=5.33cm,height=5.3cm]{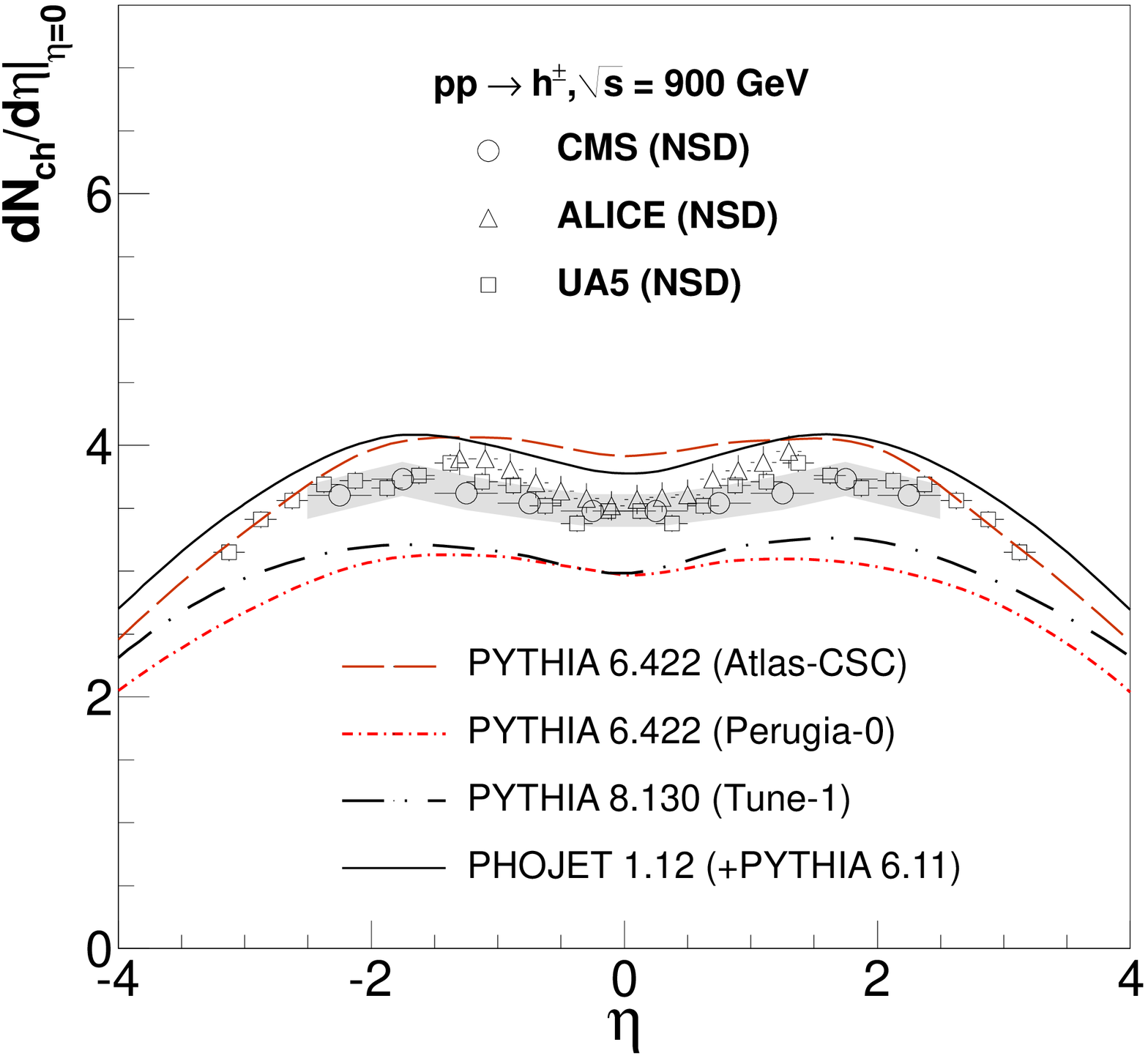}
\includegraphics[width=5.33cm,height=5.3cm]{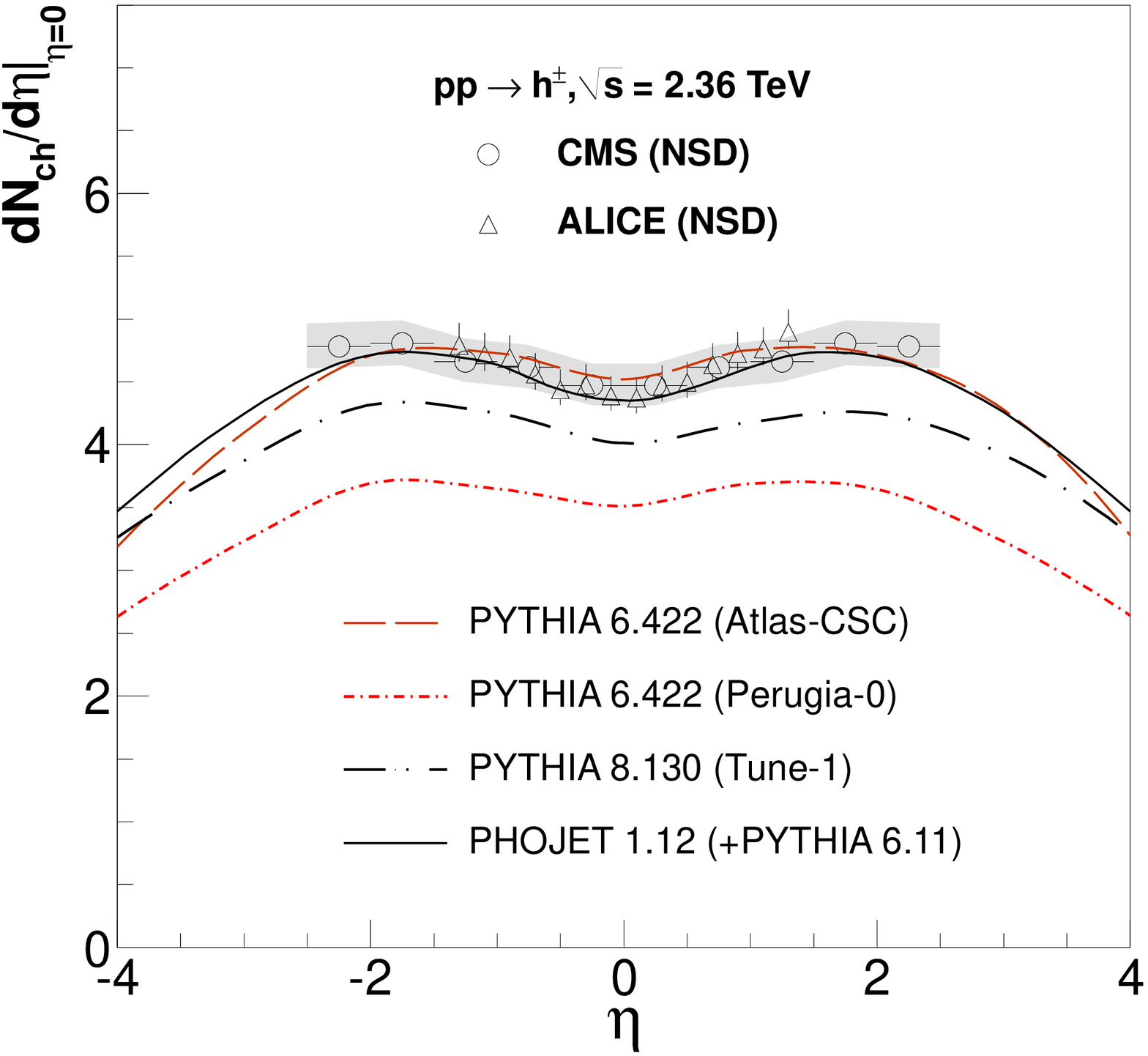}
\includegraphics[width=5.33cm,height=5.3cm]{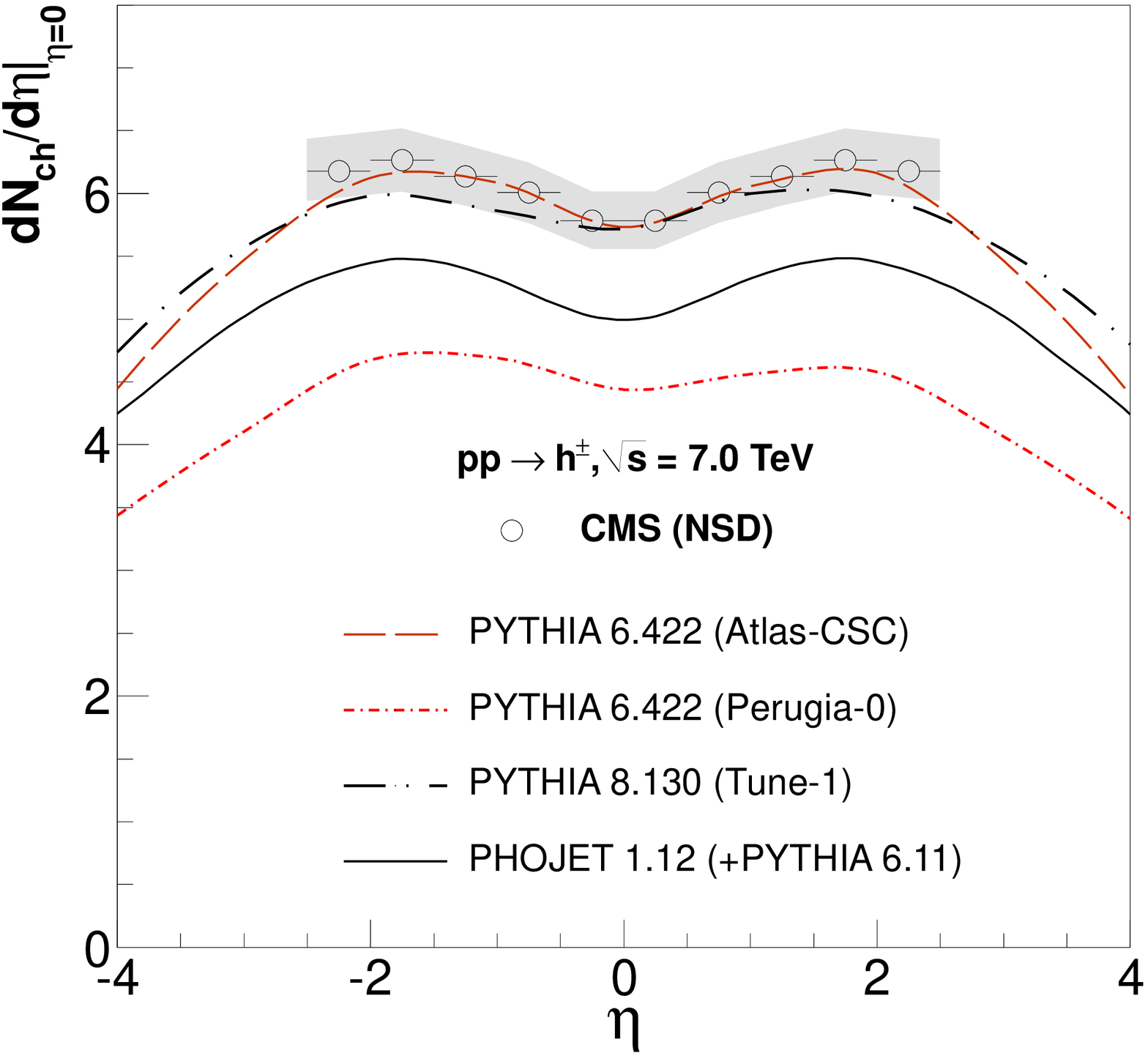}
\caption{Pseudorapidity distributions of charged hadrons, $h^\pm \equiv (h^++h^-)$, measured in 
NSD \pp\ events at the LHC ($\sqrts$~=~0.9, 2.36 and 7 TeV) by ALICE~\cite{Aamodt:2010ft,Aamodt:2010pp} and 
CMS~\cite{Khachatryan:2010xs,Khachatryan:2010us} (and by UA5~\cite{Alner:1986xu} in \ppbar\ at 900 GeV) 
compared to three different versions of \pythia\ and to the \phojet\ MC. The dashed band is the systematic 
uncertainty of the CMS experiment which is similar to those of the two other measurements.}
\label{fig:dNdeta_pythia}
\end{figure}

The pseudorapidity densities, $dN_{ch}/d\eta$, of charged hadrons measured in NSD collisions at the LHC (0.9, 2.36 and 7.0 TeV) 
by ALICE and CMS (as well as by UA5 at 900 GeV) are shown in Fig.~\ref{fig:dNdeta_pythia} compared to 
two \pythia\ 6.4 tunes, \pythia\ 8 and to \phojet. In the \pythia\ case, the NSD predictions are
obtained switching off the single-diffractive contributions\footnote{\ttt{MSUB(92)=MSUB(93)=0} in \pythia\ 6.4, 
\ttt{SoftQCD:singleDiffraction=off} in \pythia\ 8.} without any hadron-level trigger.
Since the effects of the LHC MB-selections
have been corrected for by the experiments themselves using \pythia\ (and \phojet\ as a cross-check), 
this is a consistent comparison.\\

\begin{figure}[htbp]
\includegraphics[width=5.33cm,height=5.3cm]{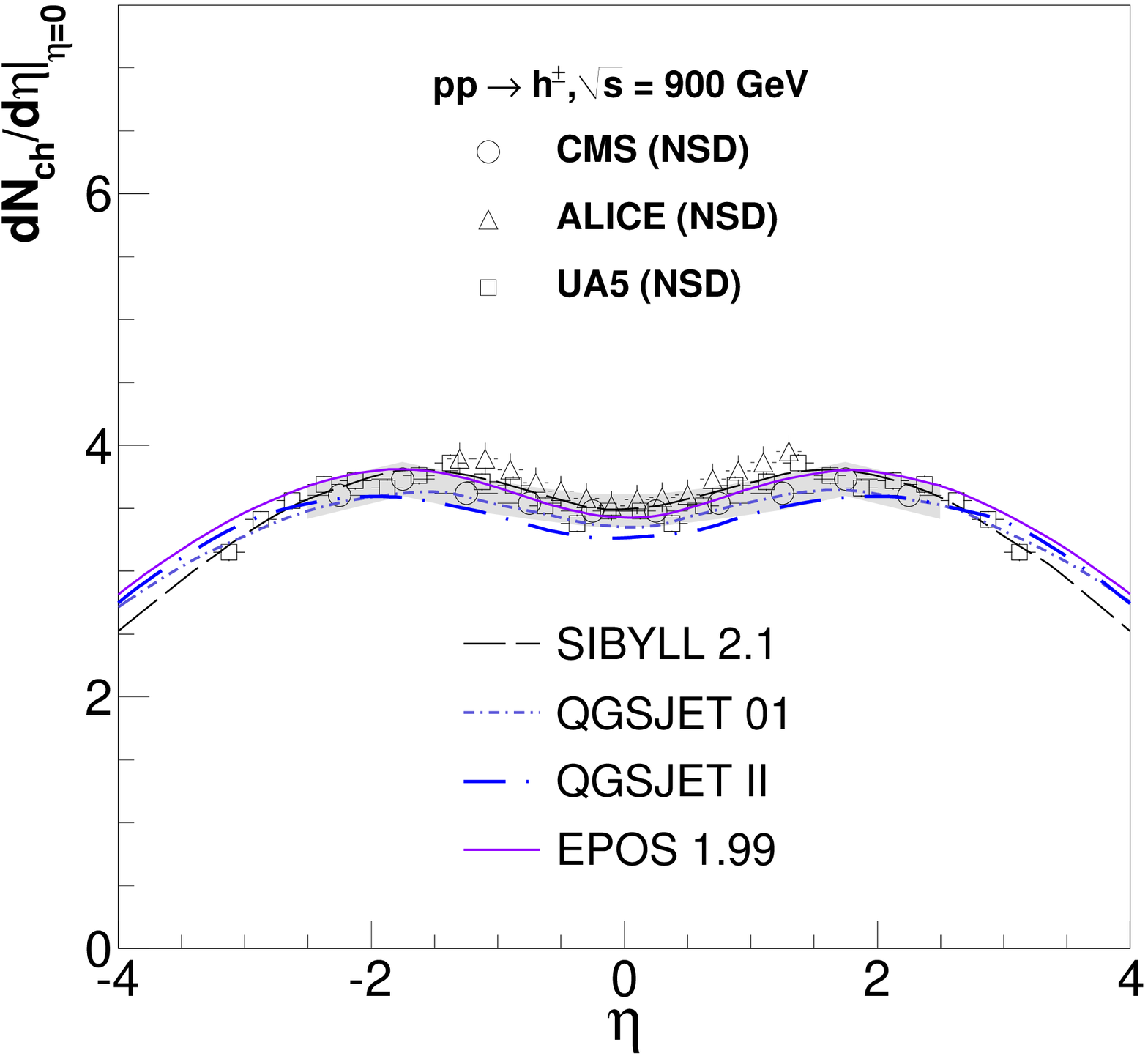}
\includegraphics[width=5.33cm,height=5.3cm]{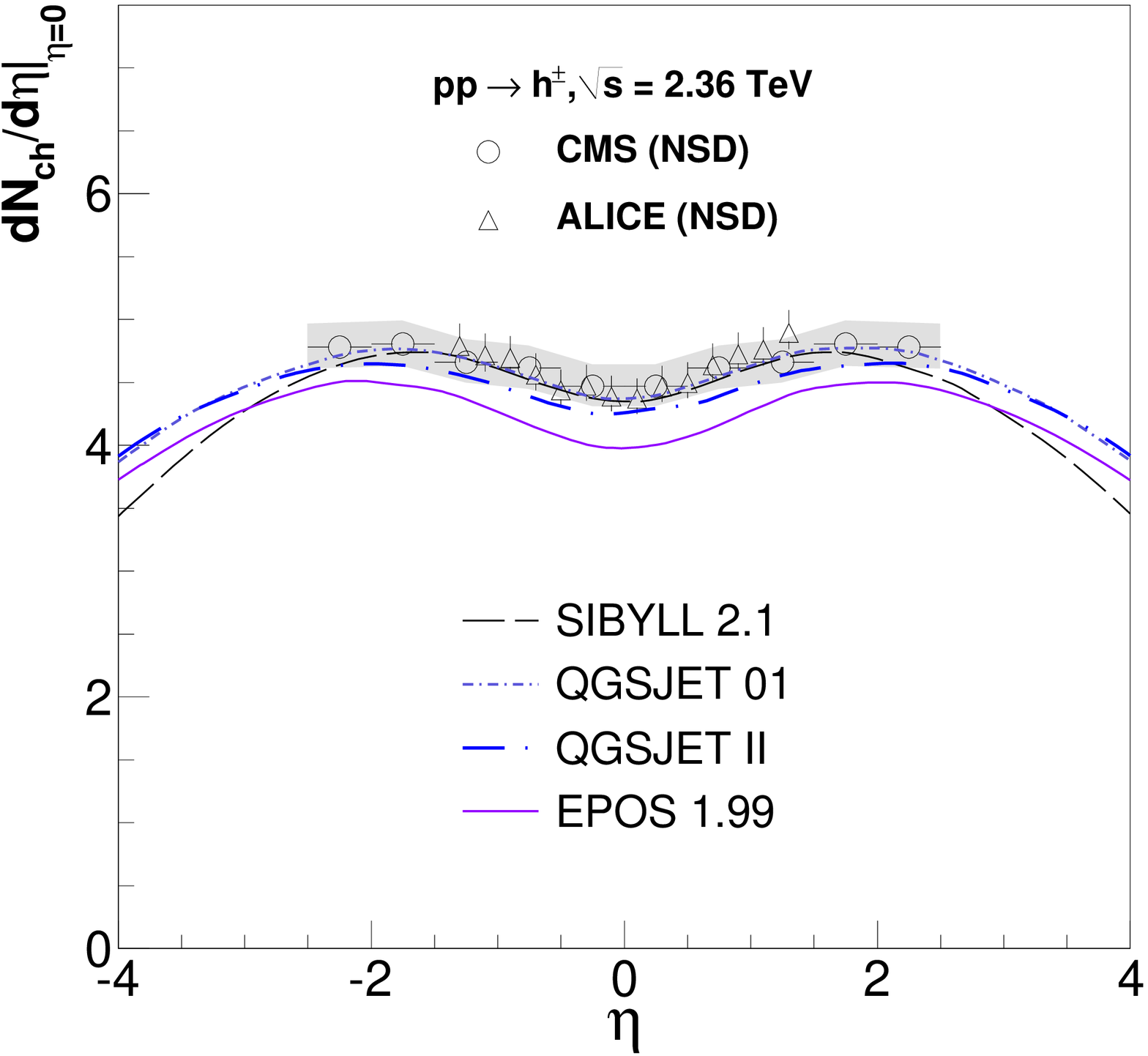}
\includegraphics[width=5.33cm,height=5.3cm]{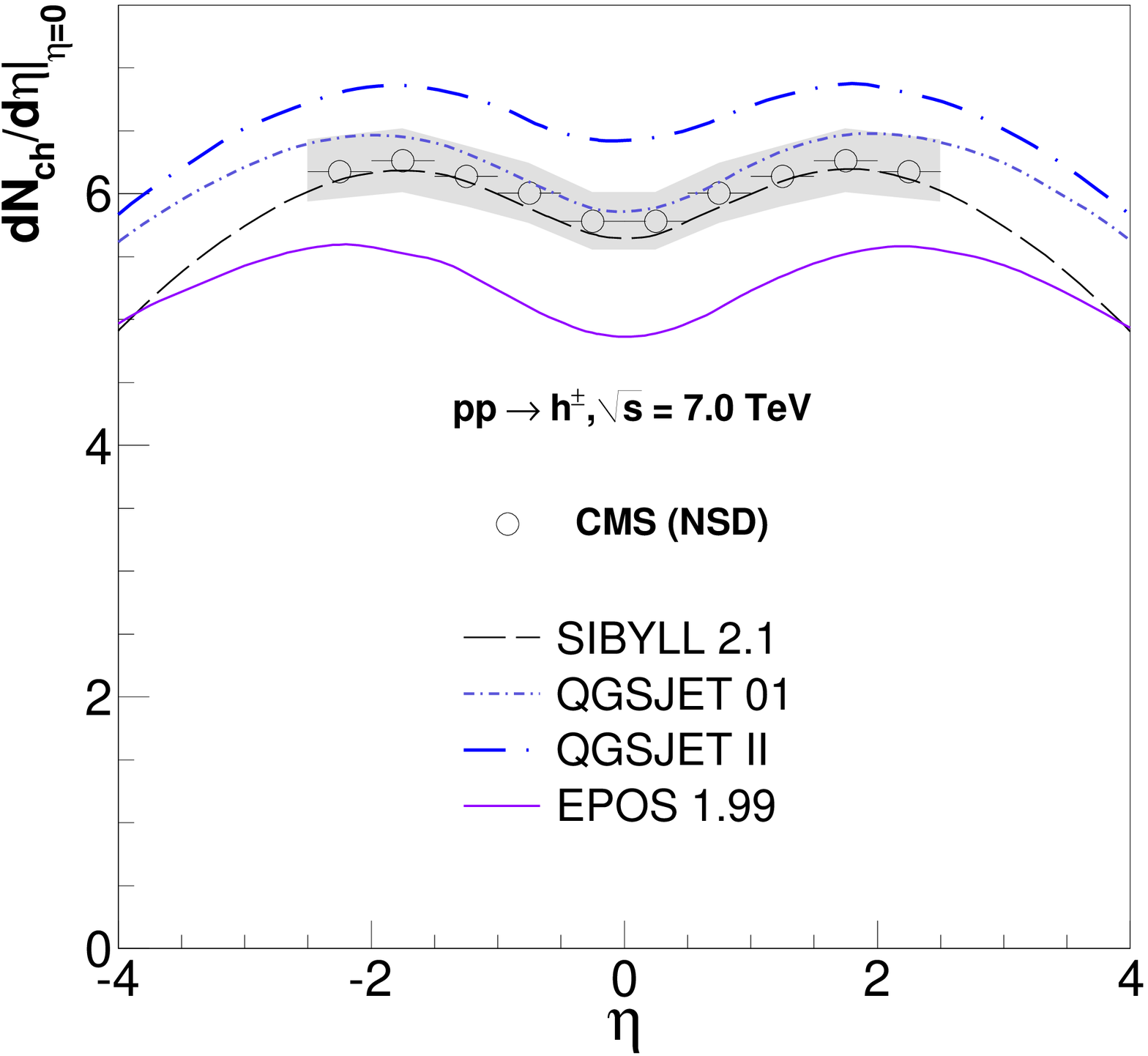}
\caption{Pseudorapidity distributions of charged hadrons, $h^\pm \equiv (h^++h^-)$, measured in NSD  
\pp\,events at the LHC (0.9, 2.36 and 7 TeV) by ALICE~\cite{Aamodt:2010ft,Aamodt:2010pp} and 
CMS~\cite{Khachatryan:2010xs,Khachatryan:2010us} (and by UA5~\cite{Alner:1986xu} in \ppbar\ at 900 GeV) 
compared to the predictions of \qgsjet 01 and II, \sibyll, and \epos. 
The dashed band is the systematic uncertainty of the CMS experiment
which is similar to those of the two other measurements.}
\label{fig:dNdeta_rft}
\end{figure}

The Perugia-0 
tune underpredicts all the measured midrapidity densities -- by about 20\% at $\sqrts$~=~0.9 and 2.36~TeV 
and by about 35\% at 7 TeV -- whereas the Atlas-CSC tune overpredicts by 10\% the central multiplicities 
at 0.9~TeV but reproduces well the data at higher c.m. energies.
\pythia\ 8 is 15\% (10\%) below the experimental central densities at $\sqrts$~=~0.9 and 2.36~TeV but 
agrees well with the experimental shape and normalization at 7 TeV. \phojet\ is compatible within 
uncertainties with the measured hadron $dN_{ch}/d\eta$ distributions at 0.9 and 2.36 TeV, whereas at 7 TeV 
it is some 15\% lower. The Perugia tunes were obtained mostly with Tevatron non-diffractive 
processes using hadrons with $\pT>$~0.4~GeV/c, which affects their prediction accuracy for the lowest 
multiplicities and lower transverse momenta considered here (the ALICE and CMS experimental hadron distributions are 
measured from $\pT\approx$~100~MeV/c and extrapolated down to zero $\pT$). The better agreement of the 
Atlas-CSC tune and \pythia\ 8 with the observations is linked to the faster rise of particle production 
with $\sqrts$, as given by their 
smaller exponents, $\varepsilon = 0.11$ and 0.08 respectively, see Eq.~(\ref{eq:12}) and Table~\ref{tab:pythiaTunes}.
The differences among the tunes cannot be related to the treatment of 
final-state effects since all of them 
have hadronization parameters tuned to LEP data and a ``weak'' option for 
the FS colour reconnection mechanism, 
which thus does {\it not} further reduce the particle multiplicity.\\

The comparison of the various cosmic-ray interaction generators to the 
NSD charged particle densities
at the LHC is shown in Figure~\ref{fig:dNdeta_rft}. 
In general the data-theory consistency is better than for \pythia.
At 900 GeV, all the RFT models are in a reasonable agreement with the measurements, 
because inclusive UA5 data were used for the model calibrations.
With increasing energy, the differences among the models predictions
increase, bracketing the data with a very good agreement for \sibyll\ and \qgsjet 01, 
and about +10\% for \qgsjet II and -20\% for \epos. Such a RFT-model-data comparison 
is not fully self-consistent because the NSD event-selection corrections applied by 
the LHC experiments 
have been obtained using \pythia\ and \phojet. We have carried out the same comparison but now
applying the CMS NSD trigger (Table~\ref{tab:triggers}) directly at the hadron-level in the simulations. 
In Table~\ref{tab:NSD-corr} we list the predicted fractions of single diffraction events, 
the efficiencies of the SD and NSD event selection, 
and the resulting corrections 
\begin{equation}
{\rm C}=\frac{dN_{{\rm ch}}}{d\eta}^{{\rm NSD-theor}}\left/\frac{dN_{{\rm ch}}}{d\eta}^{{\rm NSD-CMS}}\right.,
\label{eq:C_th_nsdcms}
\end{equation} 
for all the models considered and for different LHC energies.
The results obtained with the RFT models using the theoretical and experimental triggers
are practically identical, whereas \pythia- and \phojet-based corrections 
for the NSD event-selection resulted in about a 8\% shift down of the measured
$\dNdeta$~\cite{Khachatryan:2010xs,Khachatryan:2010us}.\\

\begin{table}[htbp]
\begin{center}
\begin{tabular}{l|ccc|ccc|ccc|ccccc}\hline
\hspace{2.0cm}Model & & \qgsjet 01 &&& \qgsjet II  &&& \epos 1.99  &&& \sibyll 2.1 & \\ 
\hspace{1.6cm}$\sqrts$ (TeV) & 0.9 & 2.36 & 7 & 0.9 & 2.36 & 7 & 0.9 & 2.36 & 7 & 0.9 & 2.36 & 7 \\ \hline
$\sigma _{pp}^{\rm SD}/\sigma _{pp}^{\rm inel}$ & 0.18 & 0.18 & 0.18 & 0.23 & 0.21 & 0.19 & 0.17 & 0.17 & 0.16 & 0.23 & 0.20 & 0.17\\
$\varepsilon _{\rm SD}$ & 6.4\% & 6.0\% & 5.0\% & 16\% & 19\% & 21\% & 29\% & 32\% & 32\% & 16\% & 22\% & 25\%\\
$\varepsilon _{\rm NSD}$ & 93\%  & 95\% & 96\% & 90\%  & 93\% & 95\% & 90\% & 92\% & 94\% & 97\% & 98\% & 99\%\\
 C & 0.99 & 0.99 & 1.0 & 0.99 & 1.0 & 1.01 & 0.97 & 0.98 & 0.97 & 1.0 & 1.02 & 1.02\\
 \hline
\end{tabular}
\caption{Fractions of single diffractive (SD) events, efficiencies of the SD ($\varepsilon_{\rm SD}$) 
and NSD ($\varepsilon_{\rm NSD}$) event-selections, and correction factors 
C, Eq.~(\ref{eq:C_th_nsdcms}), obtained applying the CMS NSD trigger 
at the hadron-level in the RFT models.} 
\label{tab:NSD-corr}
\end{center}
\end{table}

\begin{figure}[htbp]
\includegraphics[width=8.cm]{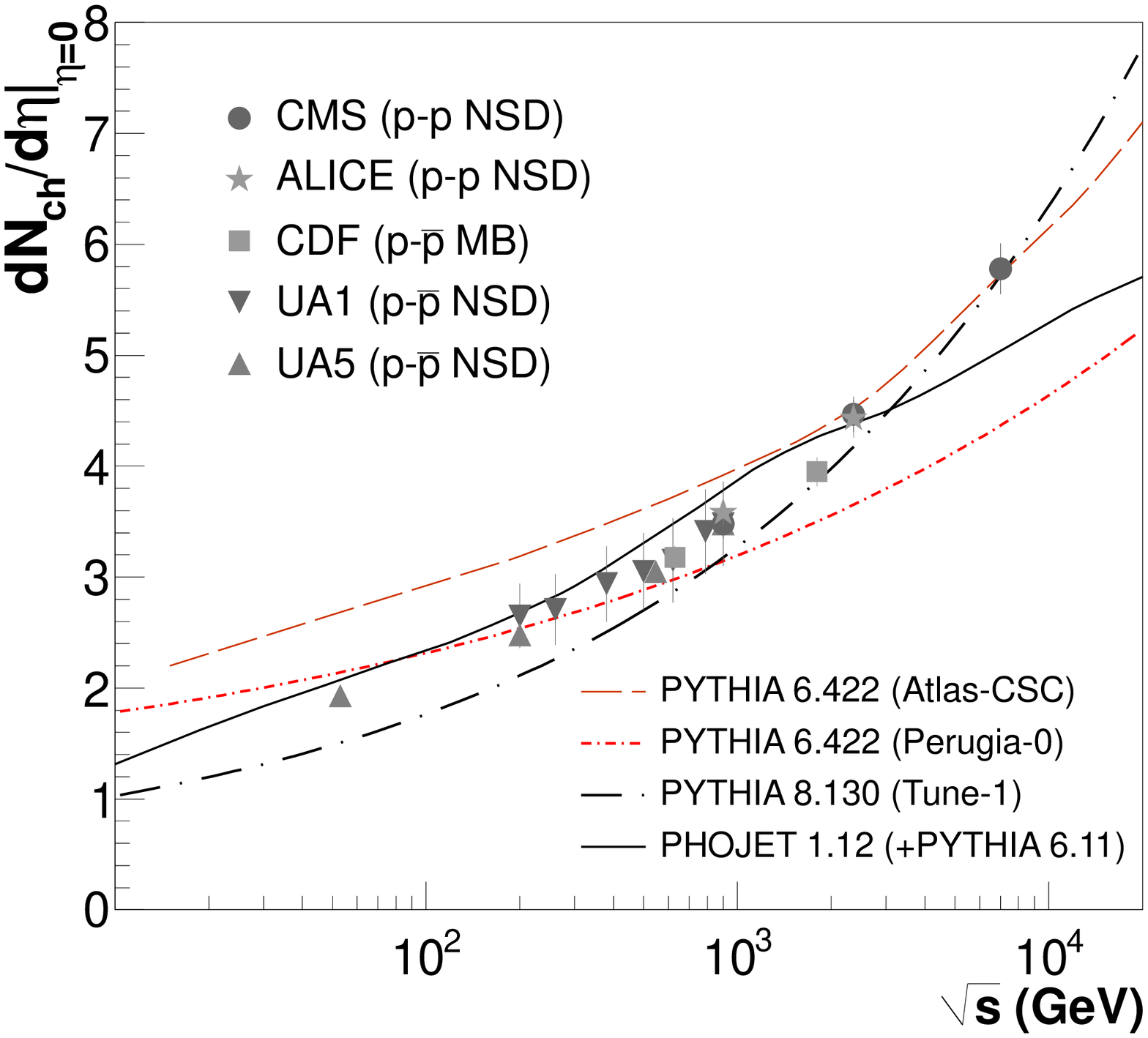}
\includegraphics[width=8.cm]{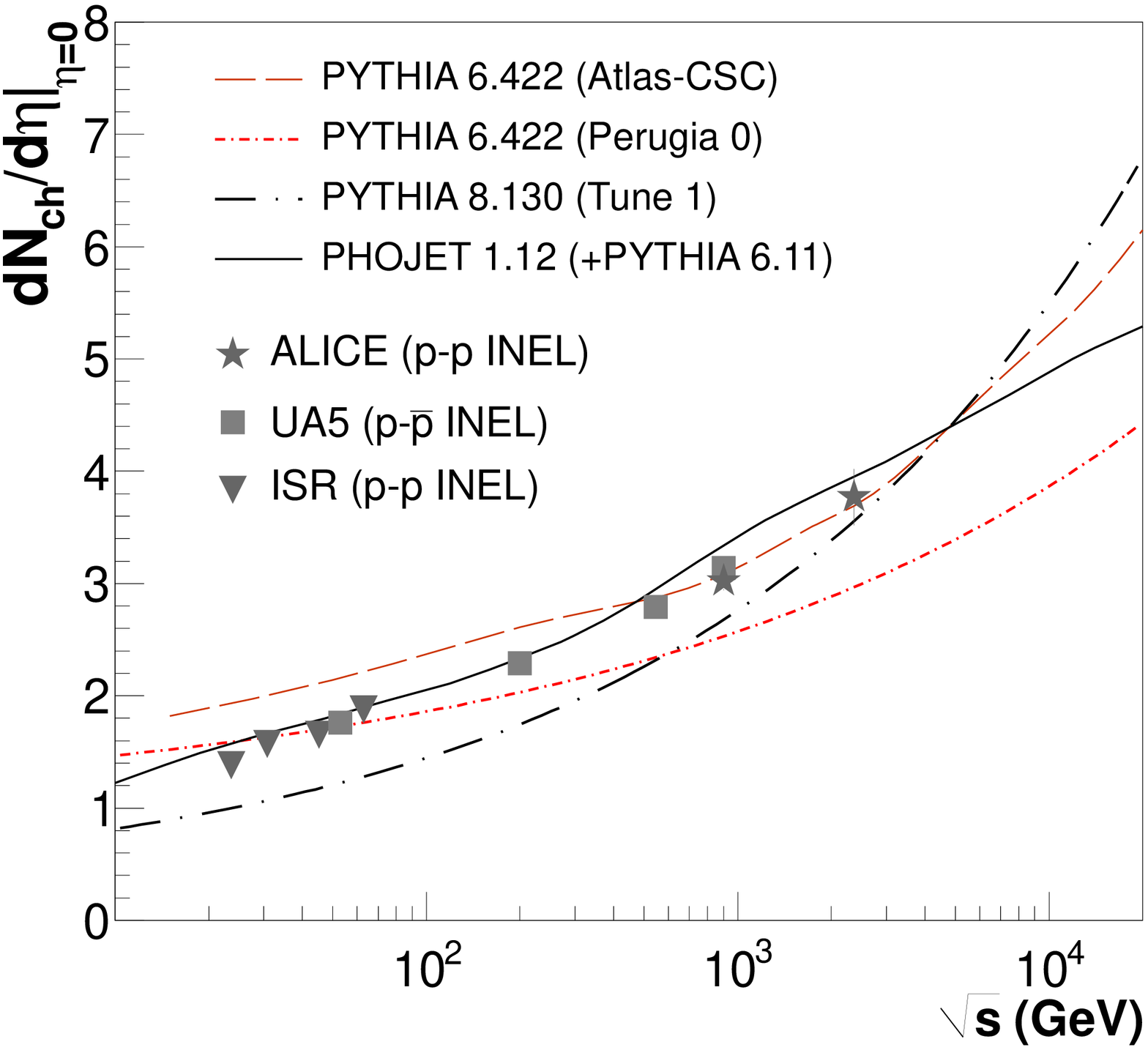}
\caption{Collision-energy dependence of the midrapidity charged hadron invariant yields in 
 non single-diffractive (NSD, left panel) and inelastic (right panel) \pp\ and \ppbar\ collisions compared 
 to different tunes of \pythia\ 6 and 8 and to \phojet\ 1.12.}
\label{fig:dNdeta_vs_sqrts_pythia}
\end{figure}

The energy dependence of the charged hadron density at $\eta = 0$ predicted by the different \pythia\
models and by \phojet\ at c.m.\ energies from $\sqrts$~=~10~GeV to $\sqrts$~=~20~TeV is presented 
in Fig.~\ref{fig:dNdeta_vs_sqrts_pythia} compared to the existing NSD (left panel) and inelastic
(right panel) data measured at Sp$\bar{\mbox p}$S, Tevatron and LHC. We note that the NSD selection 
results in measured central densities which are about 15\% {\it larger} than those obtained with 
the less-biased INEL trigger, which has less particles produced on average as it includes (most of) 
diffractive production. We notice also that the theoretical spread is smaller for the inelastic 
event-selection than for the NSD case, which indicates the different modelling of the non-diffractive 
contributions in the various MCs. Up to $\sqrts$~=~2.36~TeV, \phojet\ gives the best overall agreement with the data. 
Perugia-0 fails to reproduce the pace of increase of the central particle densities at all the energies.
The predictions of Atlas-CSC and \pythia\ 8 globally bracket the experimental energy evolution
of the hadron multiplicities at $\eta$~=~0.\\

\begin{figure}[htbp]
\includegraphics[width=8.cm]{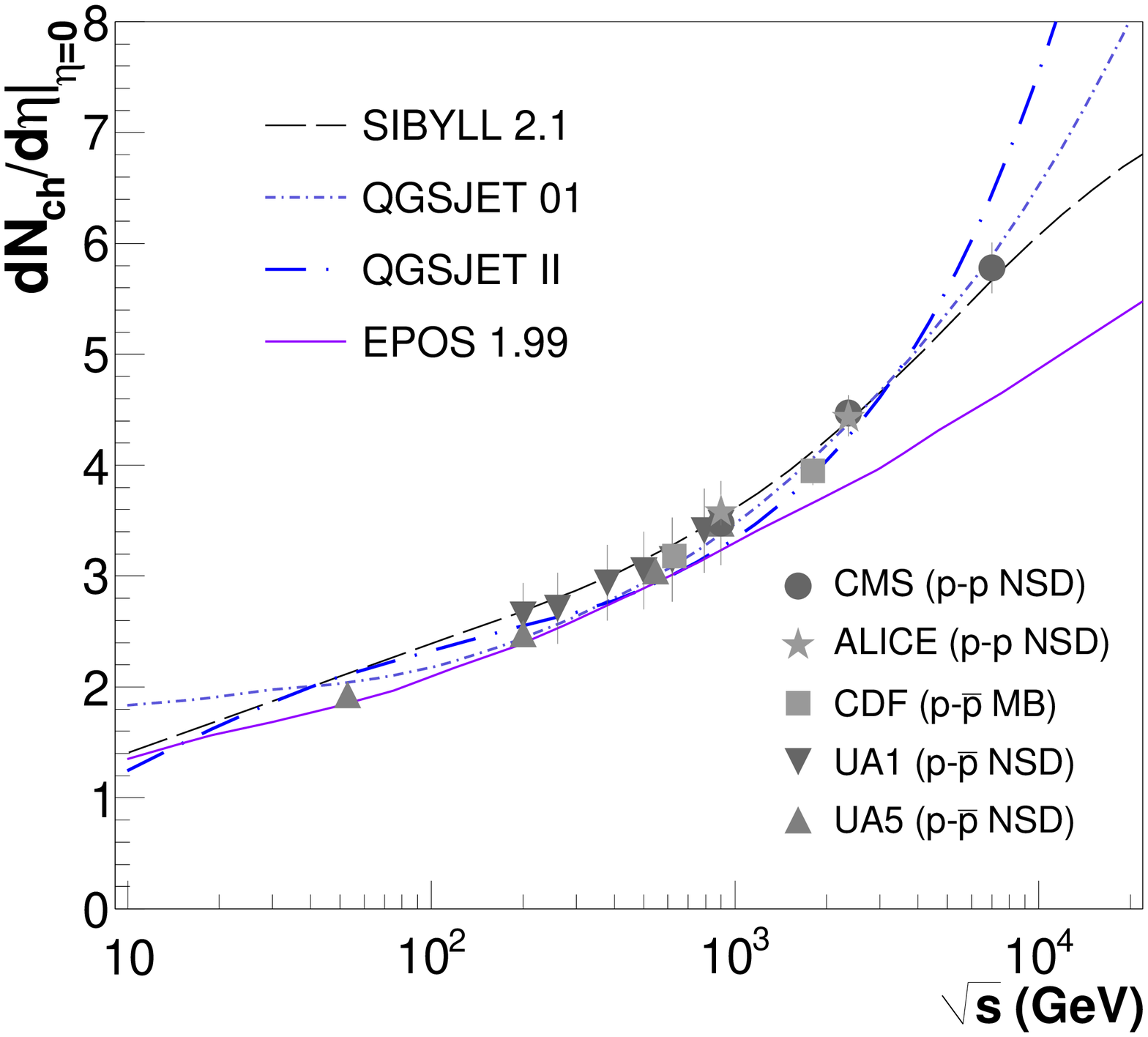}
\includegraphics[width=8.cm]{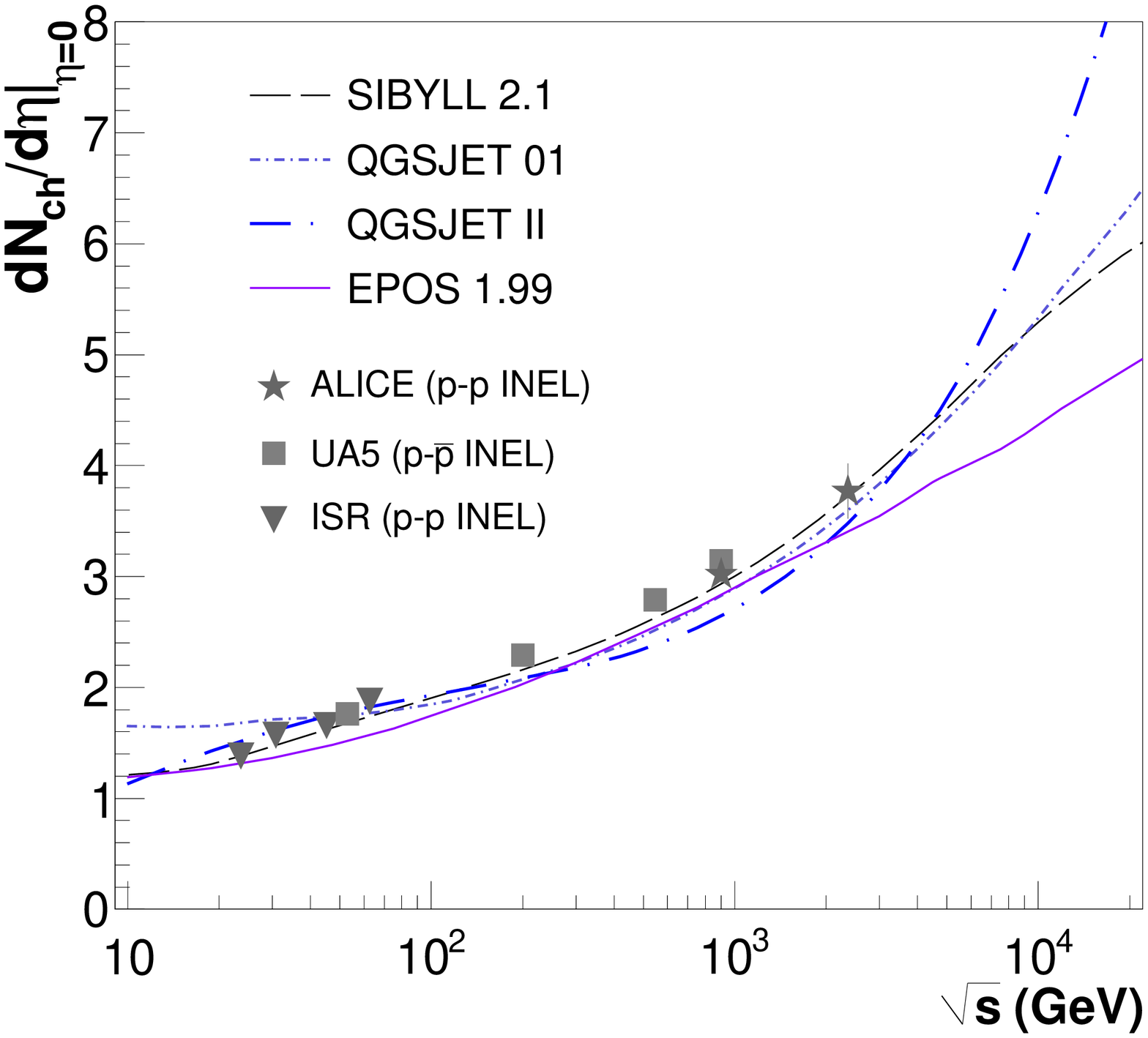}
\caption{Collision-energy dependence of the midrapidity charged hadron invariant yields in 
 non single-diffractive (NSD, left panel) and inelastic (right panel) \pp\ and \ppbar\ collisions compared 
 to the predictions of \qgsjet 01 and II, \sibyll, and \epos.}
\label{fig:dNdeta_vs_sqrts_rft}
\end{figure}

In Fig.~\ref{fig:dNdeta_vs_sqrts_rft}, the same data are compared to the RFT 
models used for air-shower simulations. The spread of model predictions at all
c.m.~energies is smaller than that of the different \pythia~tunes. Although all RFT 
models reproduce globally well the pre-LHC data up to $\sqrts\approx$~2~TeV, one 
sees different extrapolations at the current top LHC energy and beyond.
Up to 7~TeV, the older models \sibyll\ and \qgsjet 01 have better predictions 
for the average multiplicity than the newer \qgsjet II and \epos: The rate of the 
multiplicity rise between 900~GeV and 7~TeV compared to the measured one is 20\% 
higher for \qgsjet II and 15\% smaller for \epos. 

%
%

\subsection{Average transverse momentum}

The end result of a MC prediction for a given MB observable is not controlled by one single physical effect, 
but by a combination of various ingredients implemented in the model, such as e.g.~those 
listed in Tables~\ref{tab:pythiaTunes} and~\ref{tab:MCs}. For example, in \pythia\ the multiplicity density 
can be increased by allowing more underlying-event activity, and decreased by allowing e.g.\ 
more colour reconnections. Hence the same final multiplicity can be obtained through different combinations 
of MC tunings. For example, since the underlying event pumps energy into the event, in order to maintain the 
same multiplicity distribution, the hardness of the hadron spectra must then be a function of the underlying activity. 
Thus, by combining constraints from the experimental particle {\it and} transverse-momentum flows,
some extra discriminating power can be gained. As discussed in the Introduction, the mean $\pT$ in a given \pp\
event is a sensitive probe of the semi-hard dynamics (multiparton interactions and saturation effects). 
In this section, we discuss the average $\mean{\pT}$ 
predicted by the models in comparison to the collider data.\\

In the case of the \pythia\ and \phojet\ simulations we have computed $\mean{\pT}$ as done by
CMS~\cite{Khachatryan:2010xs}, i.e.~by fitting the midrapidity $\pT$-differential charged hadron spectra 
with the Tsallis function~\cite{Tsallis:1987eu}, and averaging the $\pT$ over that function.
For the RFT models we simply average the $\pT$ of all the charged particles in the central $\eta$ range.
Applying the NSD or full-inelastic selections does not change drastically the values of $\meanpt$ which
differ only by $\sim$~5\%. Also, the exact pseudorapidity coverage of the measurement around midrapidity 
(e.g.~$|\Delta\eta|<1$ or $|\Delta\eta|<2.5$) does not change much the associated mean $\pT$ values ($\sim$ -4\%) 
although an extension to full rapidities would decrease its value by about 12\%.\\


\begin{figure}[htbp]
\includegraphics[width=8.cm]{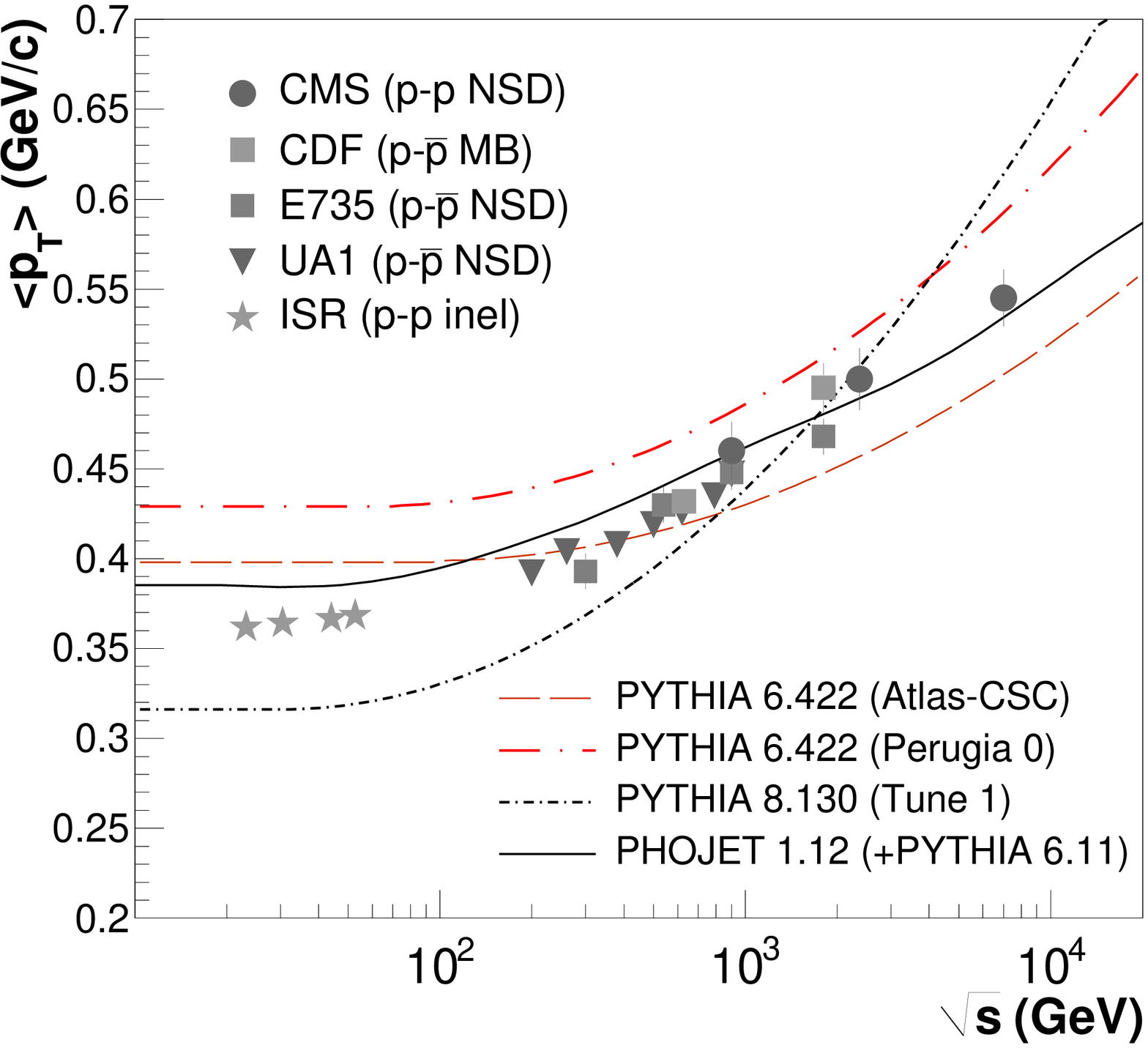}
\includegraphics[width=8.cm]{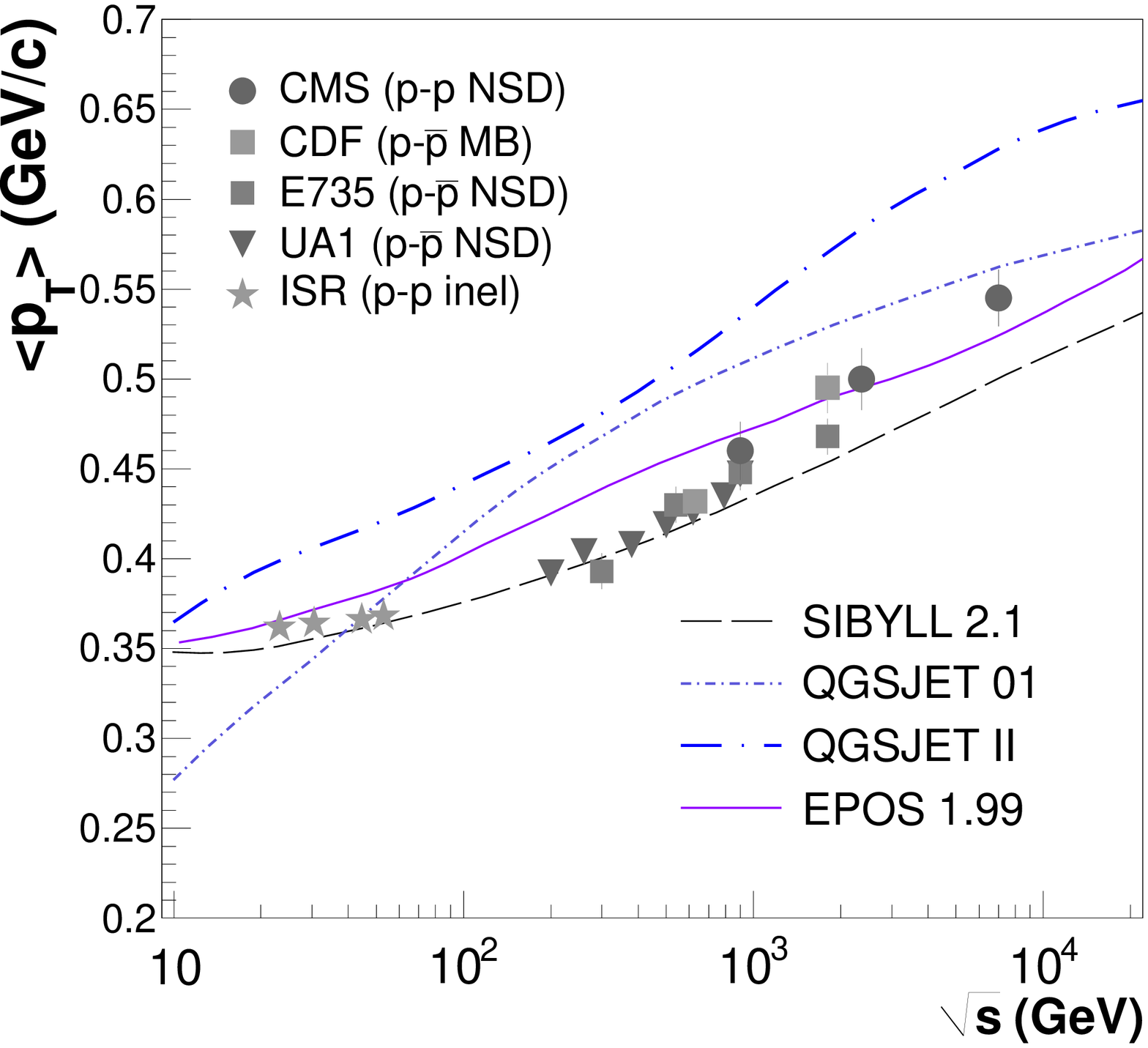}
\caption{Average $\pT$ of charged particles at midrapidity in \pp\ and \ppbar\ collisions as a function 
of $\sqrts$ compared to the \pythia\ 6 and 8 and to \phojet\ models (left panel) and to the predictions of 
\qgsjet 01 and II, \sibyll, and \epos\ (right panel).}
\label{fig:meanpT_vs_sqrts}
\end{figure}

The energy dependence of the average transverse momentum of charged hadrons measured from the ISR collider 
up to LHC energies is compared to the predictions of \pythia\ and
\phojet\ (left panel) and of cosmic ray models (right panel) in Fig.~\ref{fig:meanpT_vs_sqrts}. 
The \phojet\ and \epos\ results are globally in good 
agreement with the $\sqrts$-dependence of the average $\pT$ seen in the data. 
The Atlas-CSC \pythia\ tune and \sibyll\ predict a slower rate of increase at LHC energies.
On the contrary, the rate of the increase predicted by \pythia\ Perugia-0 and by \qgsjet II is 
compatible with the data but their absolute scale is higher by roughly 10\% and 20\% respectively. 
The \pythia\ 8 and \qgsjet 01 predictions miss the shape and absolute magnitude of $\mean{\pT}(\sqrts)$.
It is interesting to notice that the Atlas-CSC \pythia\ tune which reproduced well the pseudorapidity 
distribution  (Fig.~\ref{fig:dNdeta_pythia}) predicts a too low value for the average $\pT$, while the 
Perugia-0 tune which has a too low multiplicity shows a too large $\meanpt$.\\

%
%

\subsection{Multiplicity probability distributions}

\begin{figure}[htbp]
\includegraphics[width=8.cm,height=8.75cm]{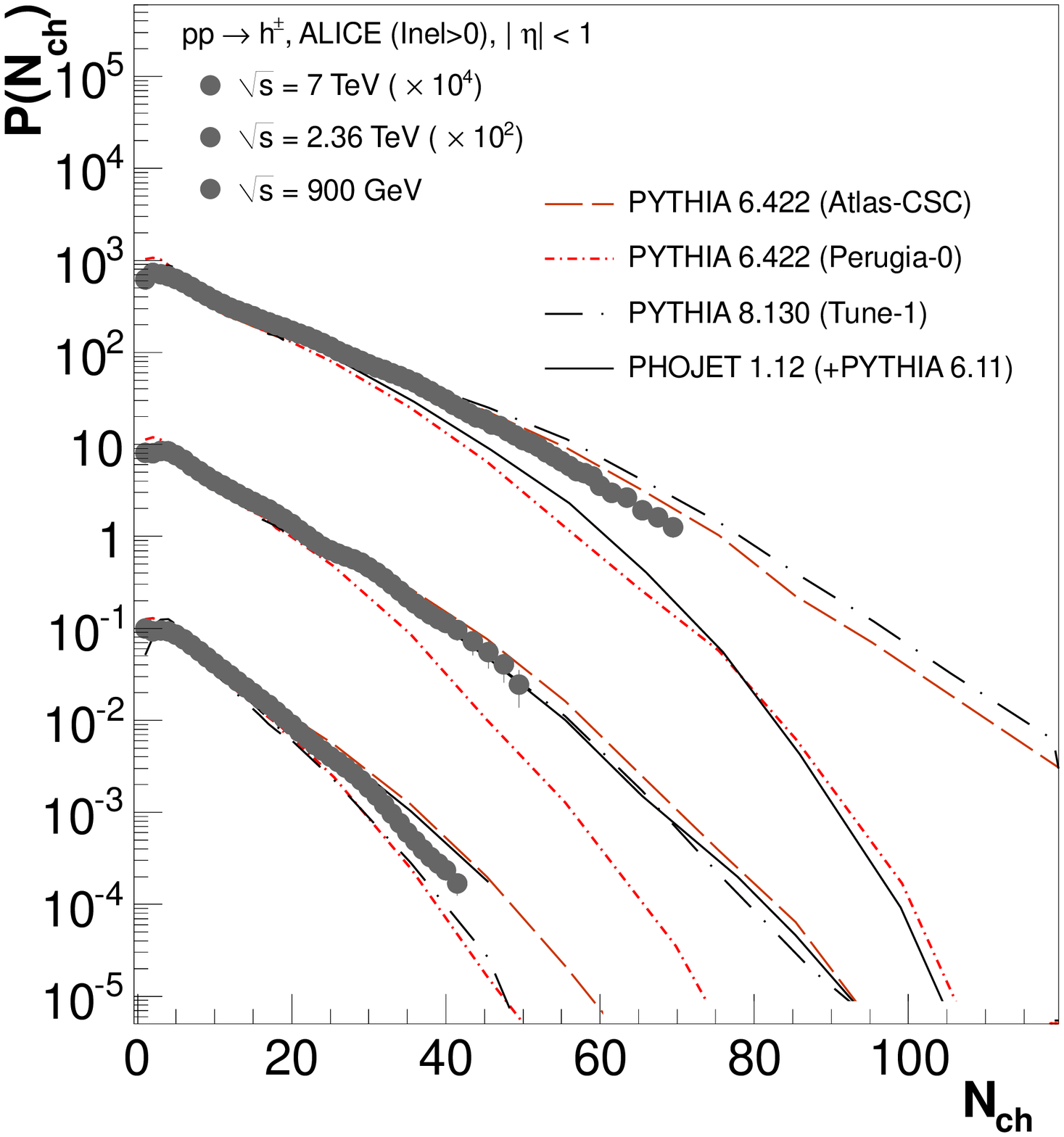}
\includegraphics[width=8.cm,height=8.75cm]{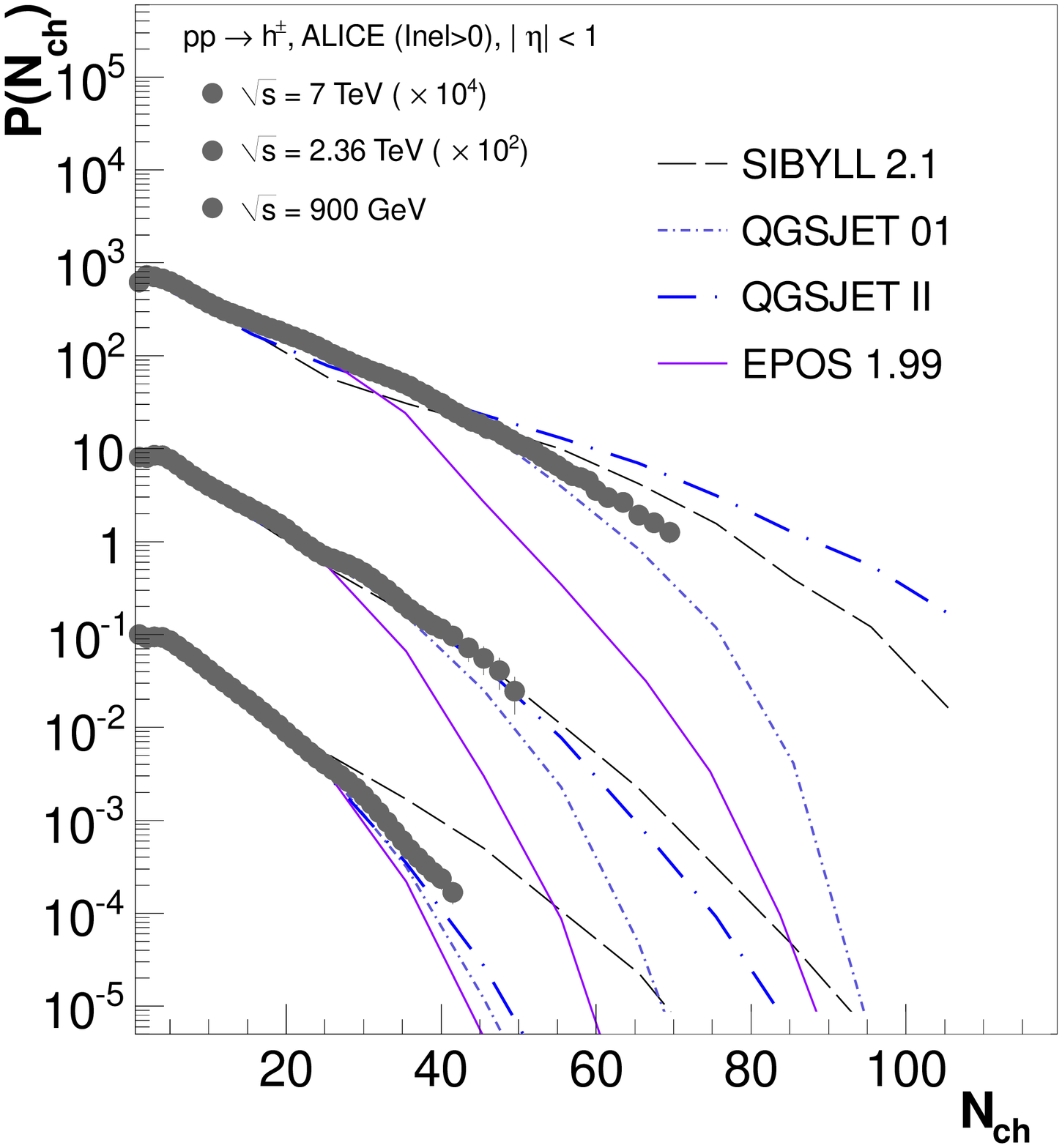}
\caption{Multiplicity distributions of charged hadrons, $P(N_{ch})$, measured by ALICE
in Inel$>$0 \pp\ events at 0.9, 2.36 and 7 TeV~\cite{Aamodt:2010pp} compared 
to the predictions of \pythia\ 6 and 8 and of \phojet\ (left plot) and of \qgsjet 01 and 
II, \sibyll, and \epos\ (right plot).}
\label{fig:mult_inel_vs_MC}
\end{figure}

\begin{figure}[htbp]
\includegraphics[width=5.25cm,height=5.cm]{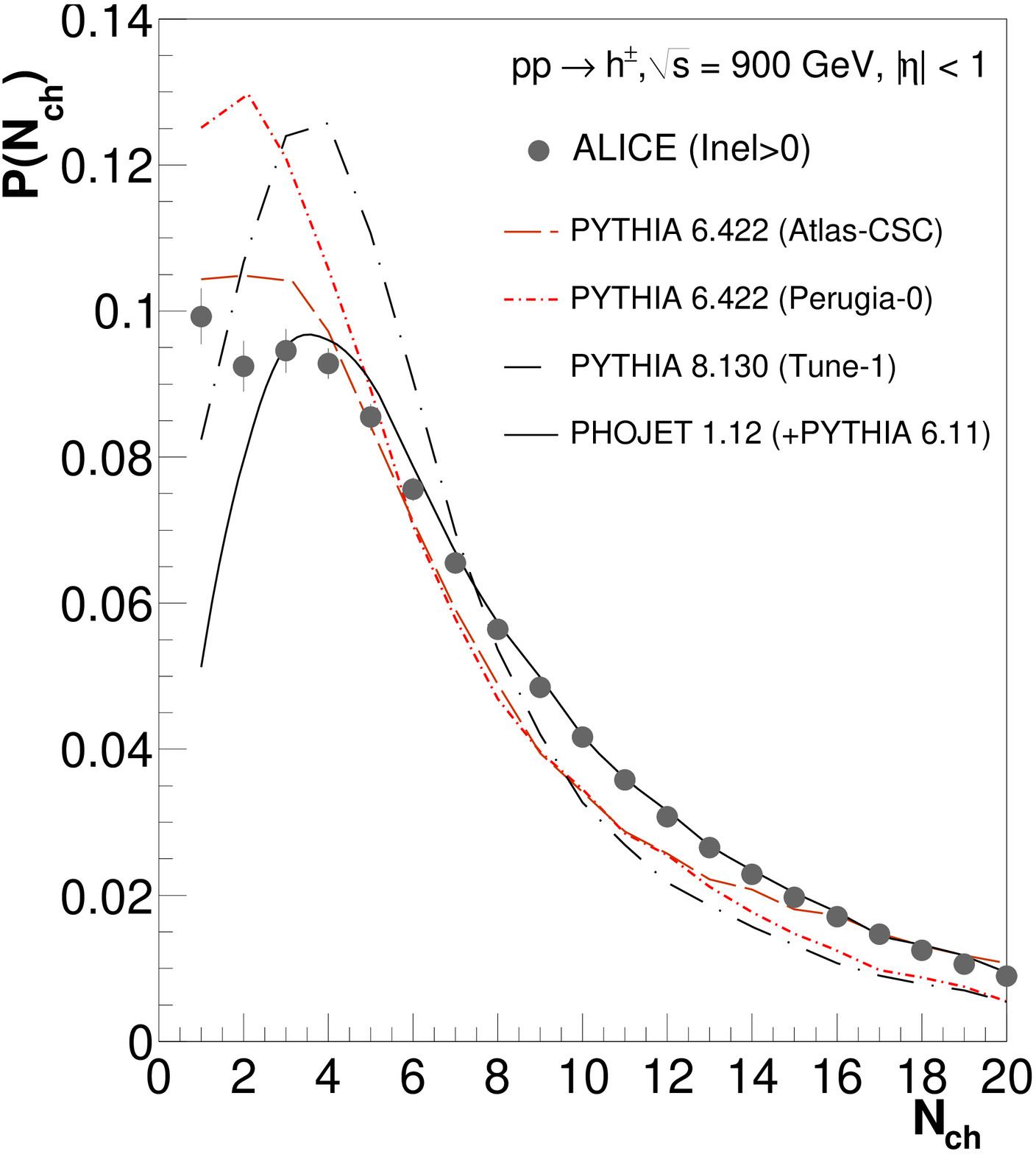}
\includegraphics[width=5.25cm,height=5.cm]{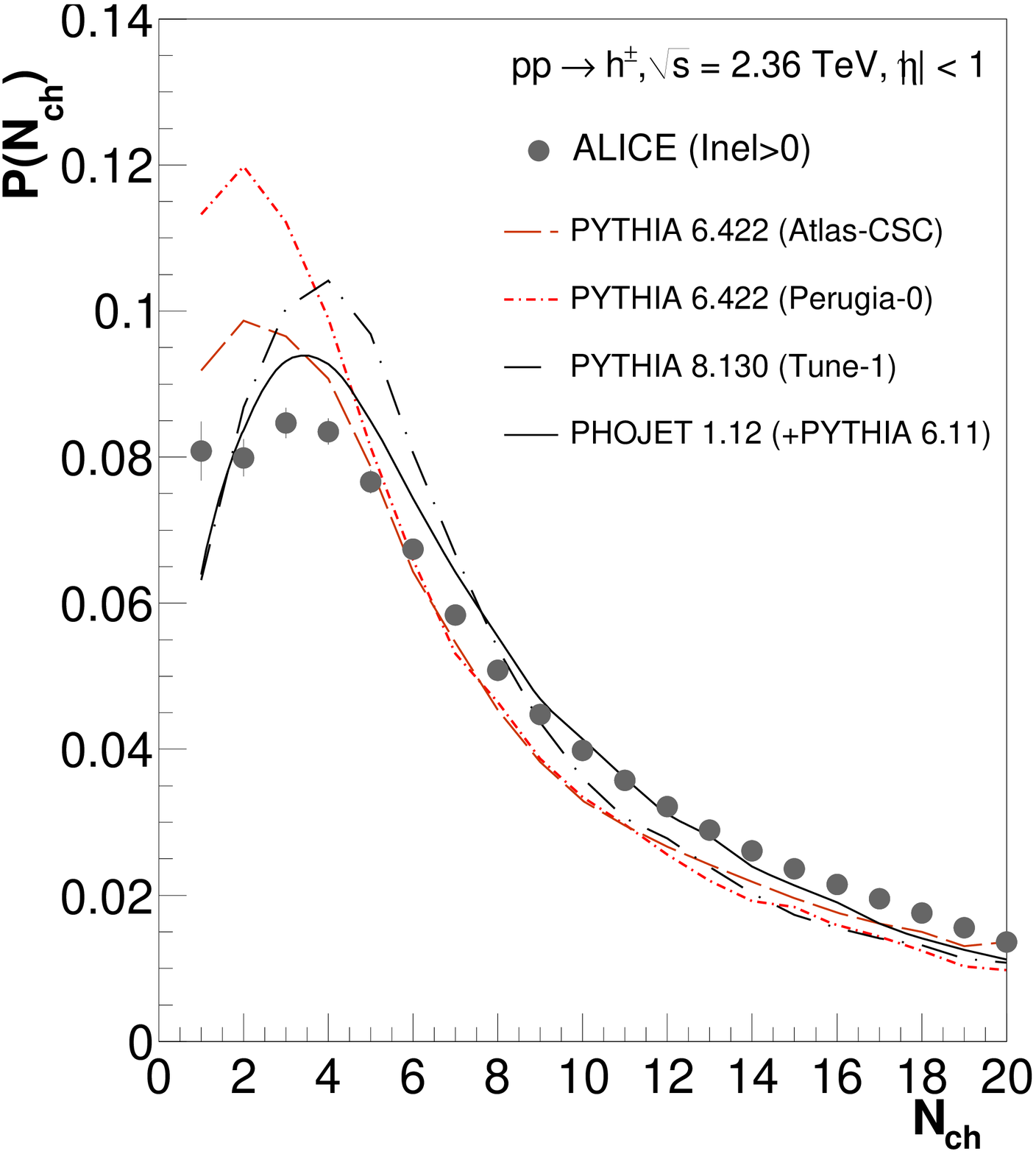}
\includegraphics[width=5.25cm,height=5.cm]{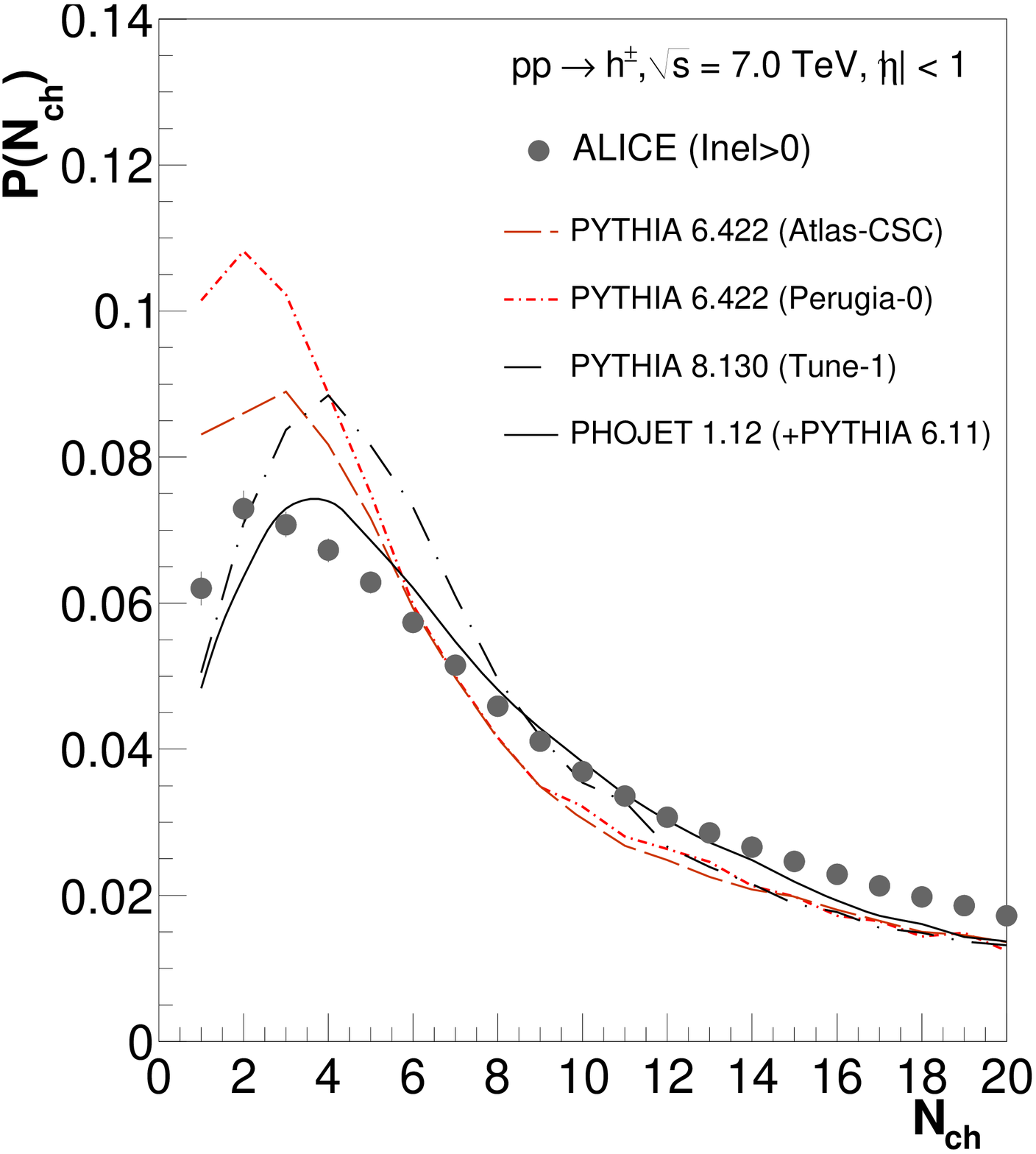}\\
\includegraphics[width=5.25cm,height=5.cm]{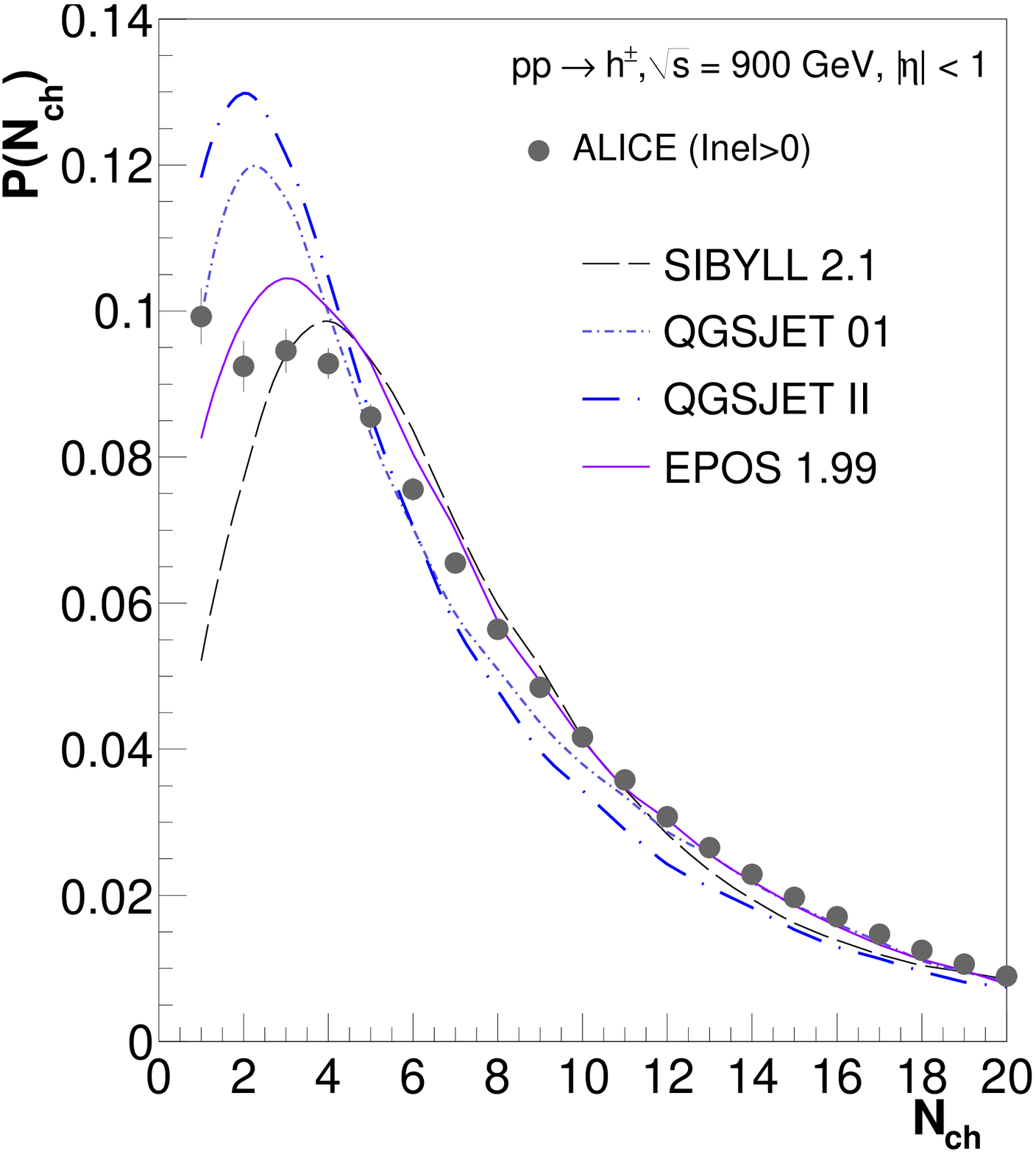}
\includegraphics[width=5.25cm,height=5.cm]{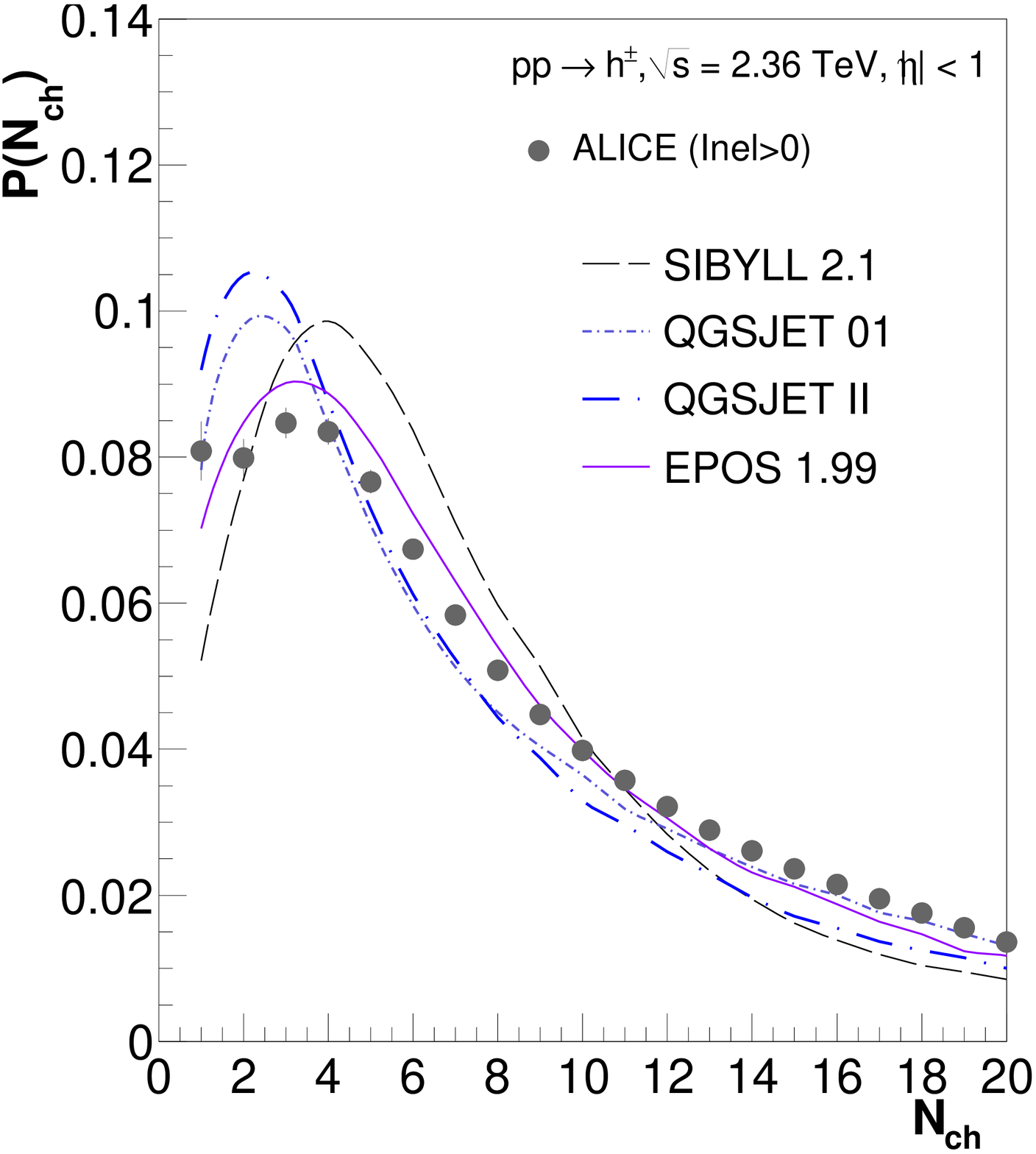}
\includegraphics[width=5.25cm,height=5.cm]{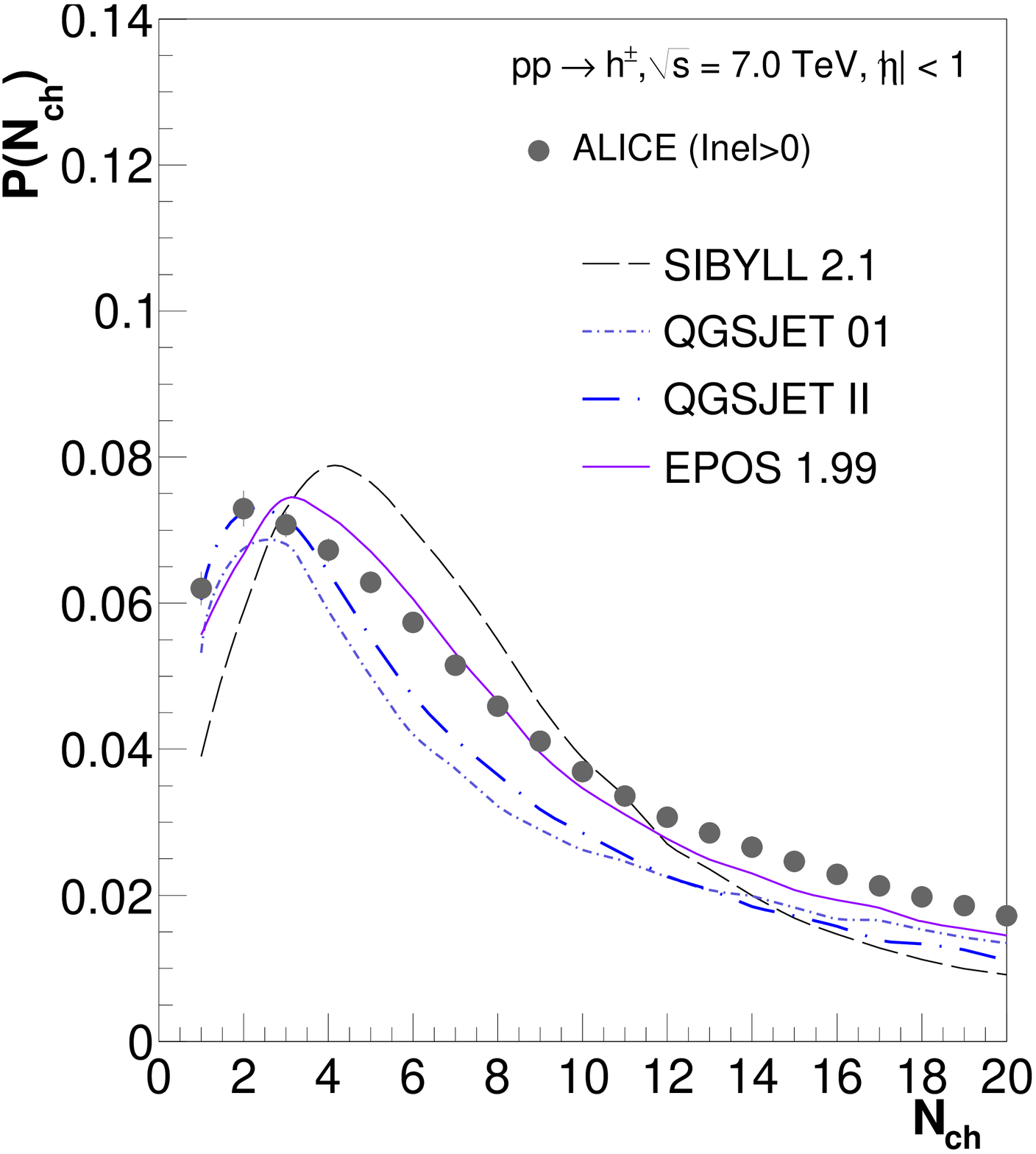}
\caption{Low part of the multiplicity distributions of charged hadrons, $P(N_{ch})$, measured 
by ALICE in Inel$>$0 events at 0.9, 2.36 and 7 TeV~\cite{Aamodt:2010pp} compared to the predictions 
of \pythia\ 6 and 8 and of \phojet\ (top panels) and of \qgsjet 01 and II, \sibyll, and \epos\ (bottom panels).}
\label{fig:mult_inel_vs_MC_zoom}
\end{figure}

The multiplicity distribution $P(N_{ch})$, i.e.~the probability to produce $N_{ch}$ charged
hadrons in an event, is of special interest because it provides extra differential 
constraints on the internal details of the hadronic interaction models. The low multiplicity part 
is mostly dominated by the contributions from 
diffraction (and from single-cut Pomeron exchanges in the RFT approaches),
whereas the tail of the distribution gives information 
on the relative contribution of multiparton scatterings (multi-Pomeron exchanges). 
The ALICE experiment has measured multiplicity distributions within $|\eta|<1$
using different triggers (inelastic, `Inel$>$0' with at least one particle measured in the 
considered $\eta$ range, and NSD) at 900~GeV, 2.36~TeV and 7~TeV~\cite{Aamodt:2010ft,Aamodt:2010pp}. 
Such different triggers affect significantly the first few bins of the distributions, where
their maxima lie. The CMS collaboration has provided a higher statistics set of results~\cite{Khachatryan:2010nk} 
but applying a NSD trigger and, thus, with large uncertainties (up to 40\%) in 
the low multiplicity part of the distributions.
In Figs.~\ref{fig:mult_inel_vs_MC} and \ref{fig:mult_inel_vs_MC_zoom}, we show the 
$P(N_{ch})$ probabilities for the ALICE `Inel$>$0' selection at the three c.m.~energies
compared to the corresponding results of \pythia, \phojet\ and RFT models. Figure~\ref{fig:mult_inel_vs_MC}
uses a log-scale to better visualize the behaviour at very large multiplicities. 
Figure~\ref{fig:mult_inel_vs_MC_zoom} shows a zoom of $P(N_{ch})$ in the low 
$N_{ch}<20$ region for \pythia\ and \phojet\ (top panels) and for the RFT models (bottom panels).\\

\pythia\ 8 and the Atlas-CSC tune of \pythia\ 6 reproduce globally well the high multiplicity tail 
at 0.9, 2.36 and 7~TeV whereas the Perugia-0 (as well as the D6T one~\cite{Aamodt:2010pp,Khachatryan:2010nk}) tune 
predicts too few hadrons at all the energies, \phojet\ is somehow in between: it reproduces the 0.9
and 2.36~TeV results but misses the tail at $\sqrts$ = 7 TeV. In the low multiplicity region,
Perugia-0 and \pythia\ 8 predict too many hadrons, whereas Atlas-CSC tune and \phojet\ are closer 
to the ALICE results. Since the modeling of diffraction is the same in Perugia-0 and Atlas-CSC,
and it is similar in \pythia\ 8 and \phojet, these results constrain well the concurrent role
of diffractive scatterings and of multi-parton interactions implemented in all these MCs.\\

In the case of RFT models, the high-$N_{ch}$ tail is underestimated by \epos\ and \qgsjet01, 
whereas \sibyll\ and \qgsjet II get a bit closer, sometimes overestimating the data. In the low-$N_{ch}$
region, $P(N_{ch})\sim$~4, only \epos\ globally reproduces the experimental results whereas the 
rest of the models overestimate the measurements up to +30\% for \sibyll. 
The peak is even shifted towards lower multiplicity in the case of both \qgsjet\ models. 
Thus, even if the average MB multiplicities at 0.9 and 2.36~TeV are well reproduced by most 
RFT models (Fig.~\ref{fig:dNdeta_rft}), the details of their probability distributions are missed
and indicate possible paths for the improvement of the different model ingredients.
For example, \epos\ has a pretty good description of the low multiplicity part but does not 
produce enough high multiplicity events (with more than 50 particles) compared to the data. 
This explains why the average $dN_{ch}/d\eta$ is too low for this model 
(see Fig.~\ref{fig:dNdeta_rft}) which seems to give 
a correct average multiplicity per cut-Pomeron but with too few Pomerons. 
The \qgsjet II model underestimates the data at the low multiplicity end but 
has a much longer tail. One possible reason for this mismatch is that the model
neglects hard Pomeron-Pomeron couplings, hence, can not describe a dynamical
evolution of the saturation scale beyond the fixed $Q_0$ cutoff value.
For \sibyll, 
the simulations oscillate around the data giving a correct average value but the 
shape of the multiplicity distribution is not described correctly.
So even if the average values of inclusive observables -- such as the multiplicity and
average transverse momentum -- are predicted correctly within 10\% by the RFT models,
there is still room for improvement of the microscopic dynamics of hadronic
interactions in order to 
safely extrapolate the predictions up to the highest cosmic-ray energies.

\clearpage

%
%

\subsection{Discussion}

The level of (dis)agreement between the three LHC inclusive hadron observables discussed 
in the previous sections and each one of the eight hadronic MCs considered in this work,
is summarized in Tables~\ref{tab:pythiaphojet_summary} and~\ref{tab:rft_summary}. As a general
conclusion, none of the models can reproduce completely all the sets of measurements.
However, some models need more retuning than others. For example, the tune Perugia-0 of \pythia~6 
does a poor job reproducing the minimum bias results measured at the LHC. As discussed previously, 
this is mainly due to the fact that the Perugia tunes were obtained mostly with Tevatron non-diffractive 
processes using hadrons with larger transverse momenta ($\pT>$~0.4~GeV/c) than those considered 
at the LHC. \pythia\ 8 has a better description of the LHC data, thanks mostly to an improved
implementation of diffractive scattering, but the default settings under Tune-1 are just 
a {\it bona fide} first guess. The Atlas-CSC tune does a better job in general but clearly cannot
reproduce diffractive-enhanced data samples measured at the LHC (not shown here, see~\cite{cms_diffraction}), 
and thus it can be mostly considered as providing a useful reference for possible settings 
-- in particular the power-law exponent $\varepsilon$ that regulates the c.m.~energy dependence of 
the infrared cutoff for (multi)parton scatterings -- that should be tried with the more 
advanced \pythia\ 8 code. As a matter of fact, recent developments in \pythia~8.145 
indicate that it is possible to reproduce LHC and previous collider data with an updated 
set of parameters (Tune 4C)~\cite{Corke:2010yf}.\\

\begin{table}[htbp]
\begin{center}
\begin{tabular}{l|ccc|ccc|ccc|ccccc}\hline
\hspace{2.2cm}\footnotesize{Model} & & \textsc{py}\footnotesize{6 Perugia0} &&& \textsc{py}6 \footnotesize{Atlas}  &&&
\pythia\,\footnotesize{8} &&& \phojet\,\footnotesize{1.12} & \\ 
\hspace{1.6cm}$\sqrts$ (TeV) & 0.9 & 2.36 & 7 & 0.9 & 2.36 & 7 & 0.9 & 2.36 & 7 & 0.9 & 2.36 & 7 \\ \hline
$\dNdeta$  & under & under & under & over & \checkmark & \checkmark & under & \checkmark & \checkmark & \checkmark & \checkmark & under\\
$\meanpt$  & \checkmark & \checkmark & over & \checkmark & under & under & \checkmark & \checkmark & over & \checkmark & \checkmark & \checkmark\\
$P(N_{ch}< 5)$ & over & over & over & \checkmark & \checkmark & over & over & over & over & under & \checkmark & \checkmark\\
$P(N_{ch}>30)$ & under & under & under & \checkmark & \checkmark & \checkmark & under & \checkmark & over & \checkmark & \checkmark & under\\
 \hline
\end{tabular}
\caption{Level of overall agreement between different \pythia\ tunes and \phojet\ 1.12
with inclusive charged hadron results measured in \pp\ collisions 
at 0.9, 2.36 and 7~TeV: pseudorapidity densities $\dNdeta$, mean transverse momentum $\meanpt$, 
and multiplicity probabilities $P(N_{ch})$ (for low and high values of $N_{ch}$). A tick (\checkmark) indicates
a reasonable data--model agreement within experimental uncertainties, and `over' (`under') 
that the MC tends to over (under) estimate the data.} 
\label{tab:pythiaphojet_summary}
\end{center}
\end{table}

The RFT models (Table~\ref{tab:rft_summary}) give an overall good description of the central 
pseudorapidity densities at LHC energies, but fail to describe consistently other
characteristics of proton-proton collisions measured by the LHC experiments.
Thus, further model improvements, particularly, those related to the treatment of 
inelastic diffraction and of the parton saturation mechanism, are highly desirable.

\begin{table}[htbp]
\begin{center}
\begin{tabular}{l|ccc|ccc|ccc|ccccc}\hline
\hspace{2.2cm}\footnotesize{Model} & & \qgsjet 01 &&& \qgsjet II  &&& \sibyll\ 2.1  &&& \epos\ 1.99 & \\ 
\hspace{1.6cm}$\sqrts$ (TeV) & 0.9 & 2.36 & 7 & 0.9 & 2.36 & 7 & 0.9 & 2.36 & 7 & 0.9 & 2.36 & 7 \\ \hline
$\dNdeta$  & \checkmark & \checkmark & \checkmark & \checkmark & \checkmark & over & \checkmark & \checkmark & \checkmark & \checkmark & under & under \\
$\meanpt$  & over & over & \checkmark & over & over & over & \checkmark & under & under & \checkmark & \checkmark & \checkmark \\
$P(N_{ch}< 5)$ & over & over & under & over & over & over & over & over & over & \checkmark & \checkmark & \checkmark \\
$P(N_{ch}>30)$ & \checkmark & under & under & \checkmark & \checkmark & over & over & \checkmark & over & under & under & under \\
 \hline
\end{tabular}
\caption{Level of overall agreement between \qgsjet 01, \qgsjet II, \sibyll\ 2.1 and \epos\ 1.99
with inclusive charged hadron results measured in \pp\ collisions 
at 0.9, 2.36 and 7~TeV: pseudorapidity densities $\dNdeta$, mean transverse momentum $\meanpt$, 
and multiplicity probabilities $P(N_{ch})$ (for low and high values of $N_{ch}$). A tick (\checkmark) indicates
a reasonable data--model agreement within experimental uncertainties, and `over' (`under') 
that the MC tends to over (under) estimate the data.} 
\label{tab:rft_summary}
\end{center}
\end{table}

%
%


\section{Implications for ultra-high energy cosmic-ray physics}
\label{sec:discuss}

\subsection{Extrapolations to ultra-high cosmic-ray energies}

In this section we show the $\sqrts$-evolution of the model
predictions for
the central rapidity density $\dNdeta$ and the average transverse momentum $\meanpt$ for  charged
hadrons produced in \pp\ collisions. We present the predicted  c.m.\ energy
dependence of these observables up to the highest cosmic-ray energies.
Both the High Resolution Fly's Eye (HiRes) experiment~\cite{Abbasi:2007sv}
and the Pierre Auger Observatory~\cite{Abraham:2008ru} 
have measured UHECR up to energies of the order of $10^{20}$~eV. Beyond this energy, the
CR flux from cosmologically distant sources ($D > 100$\,Mpc) is significantly reduced due 
to the interactions of the primary cosmic-ray particles with the cosmic microwave background 
radiation on their way from the extragalactic sources to the Earth.
This so-called Greisen-Zatsepin-Kuzmin (GZK) cutoff~\cite{Greisen:1966jv,Zatsepin66e}
corresponds to \pp\ c.m.~energies of order $\sqrt{s}_{_{\rm GZK}}\approx 300$\,TeV,
more than twenty times higher than those reachable at the CERN LHC.\\

\begin{figure}[htbp]
\includegraphics[width=8.cm]{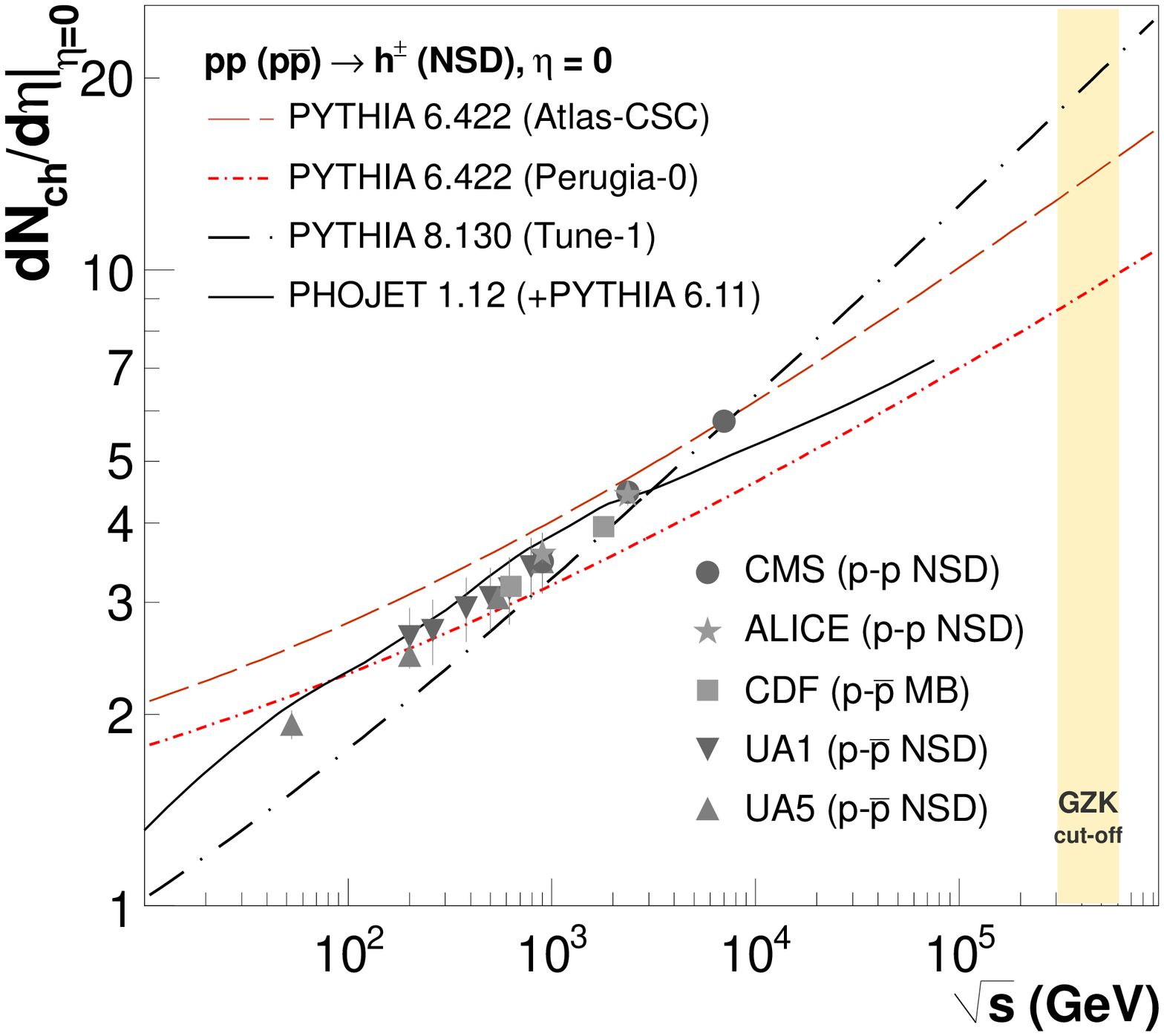}
\includegraphics[width=8.cm]{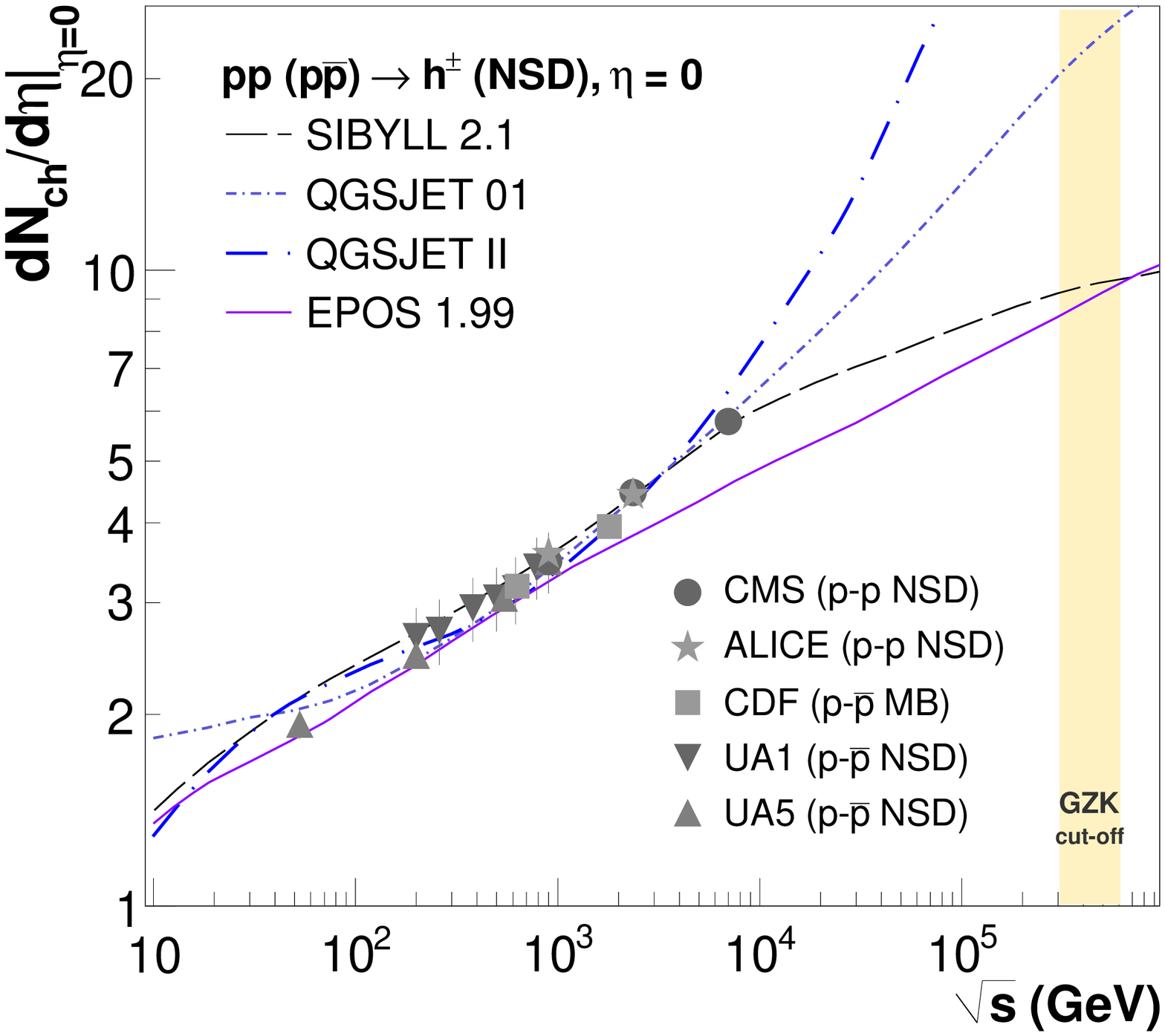}
\caption{Collision-energy dependence of the midrapidity charged hadron invariant yields in
 {\it non single-diffractive} (NSD) \pp\ collisions predicted by different tunes of \pythia\ 
 and by \phojet\ (left panel) and by \qgsjet 01 and II, \sibyll, and \epos\ (right panel)
 MCs up to the GZK cutoff energies. The data points are the same 
 as in Fig.~\ref{fig:dNdeta_vs_sqrts_pythia} (left).}
\label{fig:dNdeta_vs_sqrts_nsd_gzk}
\end{figure}

\begin{figure}[htbp]
\includegraphics[width=8.cm]{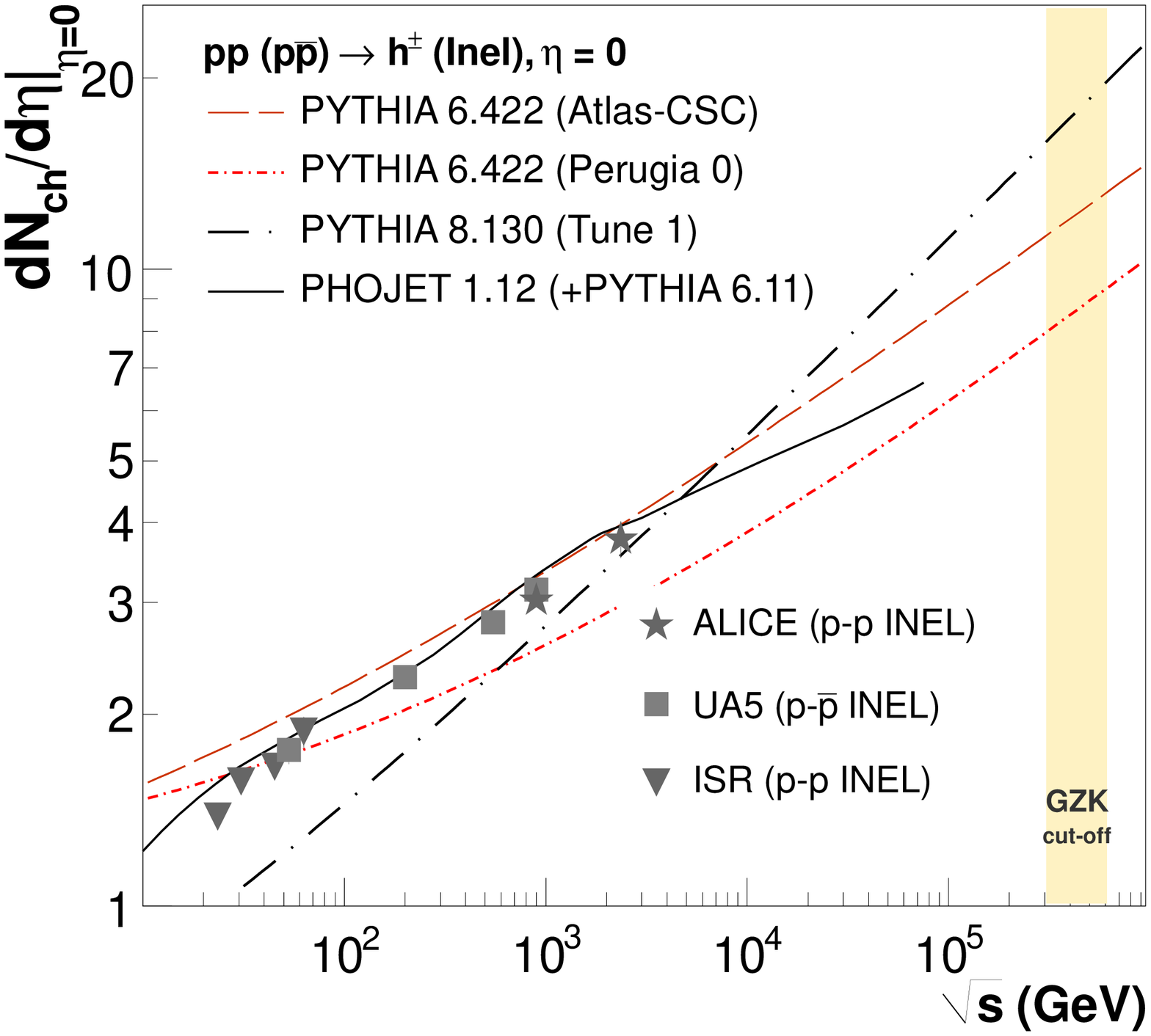}
\includegraphics[width=8.cm]{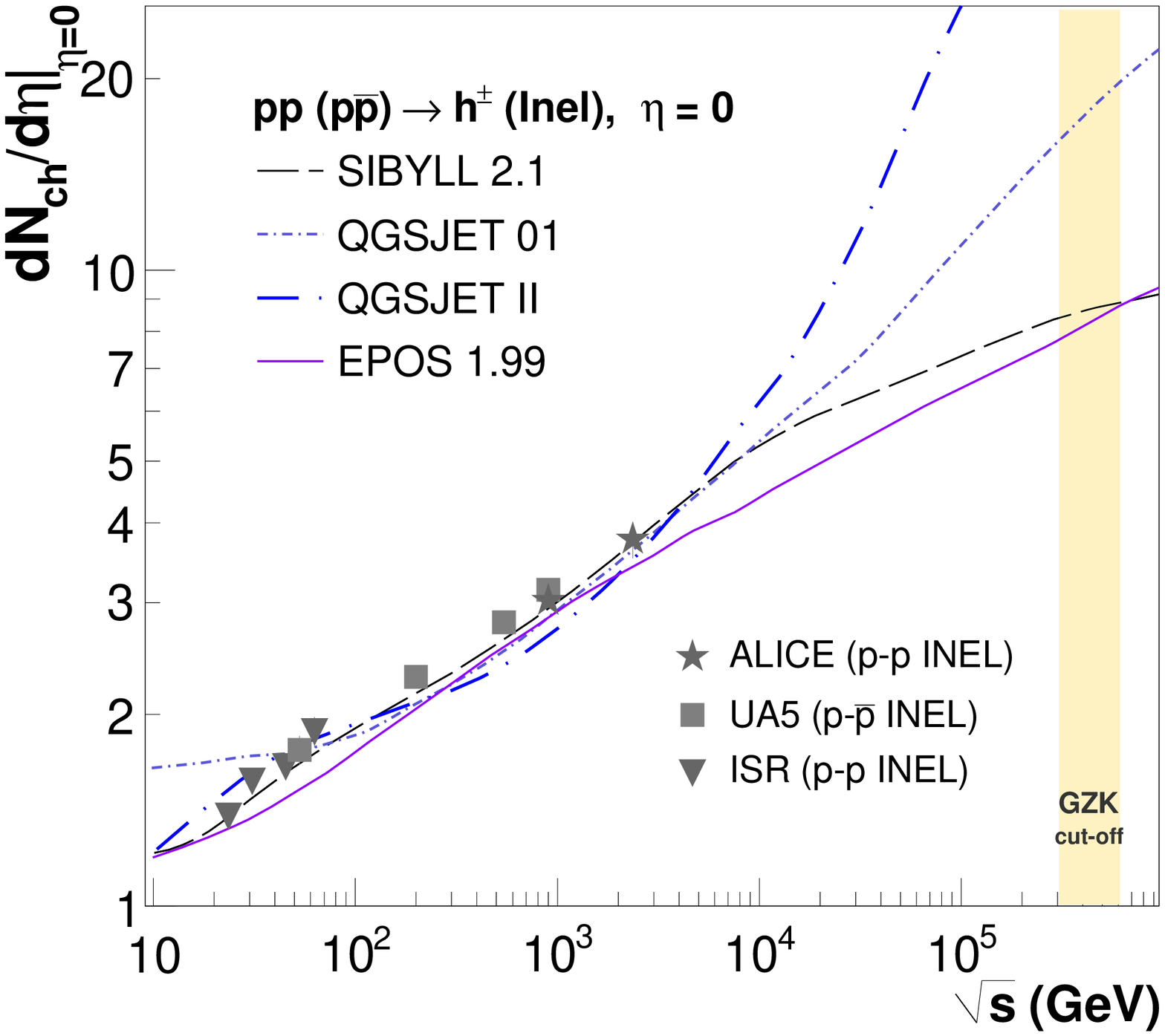}
\caption{Collision-energy dependence of the midrapidity charged hadron invariant yields in
 {\it inelastic} \pp\ collisions predicted by different tunes of \pythia\ and by 
 \phojet\ (left panel) and by \qgsjet 01 and II, \sibyll, and \epos\ (right panel) MCs up to the GZK 
 cutoff energies. The data points are the same as in Fig.~\ref{fig:dNdeta_vs_sqrts_pythia} (right).}
\label{fig:dNdeta_vs_sqrts_inel_gzk}
\end{figure}

In Figs.~\ref{fig:dNdeta_vs_sqrts_nsd_gzk} and~\ref{fig:dNdeta_vs_sqrts_inel_gzk} we present the energy 
evolution predicted by all the models for $\dNdeta$ up to GZK energies for NSD and inelastic 
\pp\ interactions respectively. The left panels show the predictions of the various \pythia\ versions
considered and of \phojet, and the right panels those of the rest of the RFT generators. 
\phojet\ predictions stop at around $\sqrts\approx$~100~TeV, which is the maximum c.m.~energy where
the model can be safely extrapolated to within its current implementation.\\

The first thing to notice is that all \pythia\ tunes as well as \phojet, \qgsjet 01 and \epos\ feature, 
with different slopes, a power-law dependence $dN/d\eta\propto (\sqrts)^{n}$ -- i.e.~a linear 
behaviour in ($\log \sqrts, \log \dNdeta$) scales -- of the midrapidity particle-density. 
On the other hand, \qgsjet II and \sibyll\ show faster and slower
dependencies on energy, respectively. Disregarding for the moment the
\pythia\ Perugia-0 and \epos\ 1.99 predictions, which are already 
lower than the particle multiplicities measured at 7 TeV, we see that the rest of the models can more or less 
reproduce the collider data up to LHC energies. Still they have very
different extrapolations at the GZK energy range.
The highest central rapidity densities at $\sqrt{s}_{_{\rm GZK}}$ are predicted by \qgsjet II which 
reaches $\dNdeta\approx$~50 (off-scale in the plot), \qgsjet 01 and \pythia\ 8 predict values of the order 
$\dNdeta\approx$~20, the Atlas-CSC \pythia\ 6.4 tune indicates $\dNdeta\approx$~14, and finally 
\sibyll\ (as \epos) a low $\dNdeta\approx$~10 value at the GZK cutoff.\\

\begin{figure}[htbp]
\includegraphics[width=8.cm]{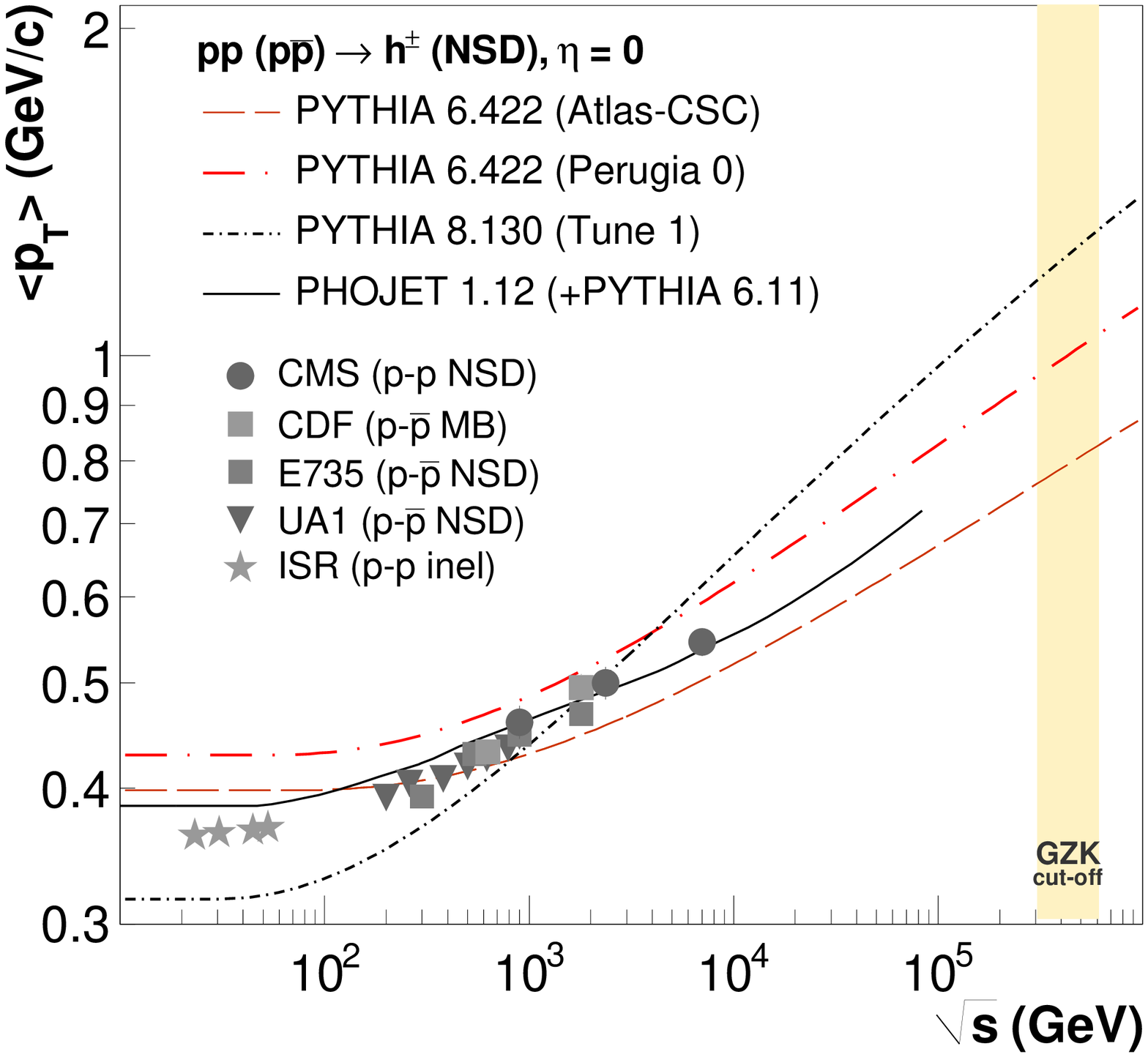}
\includegraphics[width=8.cm]{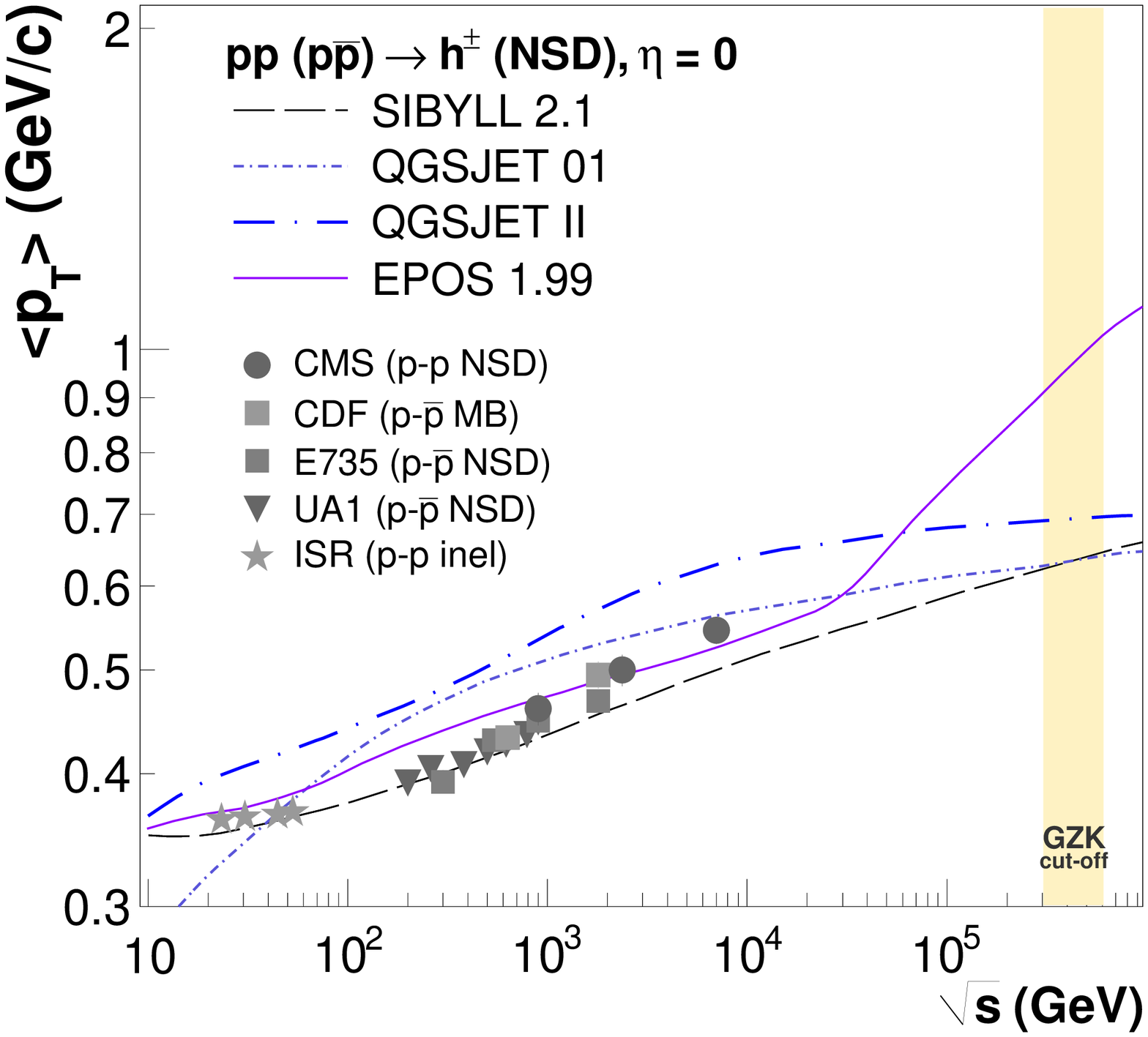}
\caption{Collision-energy dependence of the midrapidity average $\pT$ in
 {\it non single-diffractive} (NSD) \pp\ and \ppbar\ collisions predicted by different tunes 
 of \pythia\ and by \phojet\ (left panel) and by \qgsjet 01 and II, \sibyll, and \epos\ 
 (right panel) MCs up to the GZK cutoff energies. The data points are the same 
 as in Fig.~\ref{fig:meanpT_vs_sqrts}.}
\label{fig:meanpT_vs_sqrts_gzk}
\end{figure}

In Fig.~\ref{fig:meanpT_vs_sqrts_gzk} we show the energy evolution of the average $\pT$ predicted
by \pythia\ 6 and 8 and by \phojet\ (left panel) and by the cosmic-ray event
generators (right panel).
Above $\sqrts\approx$~100~GeV, all the models show a power-law behaviour (with varying exponents) 
of $\meanpt$ with collision energy. Interestingly, all the RFT MCs with the exception of \epos\ 
predict a very moderate increase of $\meanpt$ with energy: the $\meanpt$ amounts to 
$\sim$0.6~GeV/c at GZK energies which is only 0.05~GeV/c above the current CMS result at 7 TeV. 
On the other hand, \pythia\ 8, \pythia\ 6.4 (Atlas-CSC) and \epos\
indicate $\meanpt_{_{\rm GZK}}\approx$~1~GeV/c (\phojet, with an extrapolated 
$\meanpt_{_{\rm GZK}}\approx$~0.8~GeV/c, is somewhat in between).\\

The different energy behaviour of the $\meanpt$ reflects directly
the assumptions made in the models related to the low-$x$
behaviour of the PDFs. For example, in models with saturation of parton
densities, the mean transverse momentum of the produced hadrons is of
the order of the saturation scale $Q_0$ in the high-energy limit.
In the case of the two \qgsjet\ models, the $\meanpt$ does initially increase
with energy and later approaches an
asymptotic value. This behaviour is related to the fact that parton
saturation is neglected in \qgsjet 01 while in \qgsjet II there is no dynamical
evolution of the saturation scale above the fixed  $Q_0$ value.
In contrast,
assuming saturation also for highly virtual partons,  saturation
effects lead to a continuous increase of the average $\pT$ faster than 
$\log s$, as found  in \phojet\ or \pythia. \epos\ would have had a 
behaviour similar to \qgsjet, but because of the final-state collective 
parton expansion is taken into account in this model, the generated 
flow increases the average $\pT$ rapidly above $\sqrts\approx$~20~TeV.


%
%

\subsection{Implications for the interpretation of extensive air showers}


Extensive air showers initiated by interactions of primary cosmic ray
particles (protons and nuclei) with air nuclei in the upper atmosphere
constitute multi-step cascade processes. The backbone of an air shower is
the hadronic cascade of interactions of both primary and secondary hadrons
-- (anti)nucleons, pions, and kaons -- which quickly dissipates the initial 
energy between many hadronic particles and which pumps the energy into
secondary electromagnetic (e.m.) cascades, mostly via decays of the produced neutral pions.
In turn, high energy e.m. cascades proceed mainly via pair production of
electrons and positrons from photons and by $e^{\pm}$ bremsstrahlung,
with a relatively weak feedback to the hadronic cascade via photonuclear
interactions.\\

Experimental methods of EAS detection include measurement of charged particles
arriving at ground level and studies of longitudinal shower development --
mostly via observations of fluorescence and \v{C}erenkov light produced by 
charged particles ($e^{\pm}$ mainly) -- see~\cite{Haungs:2003jv} for a review 
of detection techniques. The most important shower observables are the position of
the shower maximum, $X_{\max}$, i.e.~the depth in the atmosphere (in g/cm$^2$) 
where the number of charged particles reaches its maximum, the number of particles 
at maximum $N_{\max}$, and the number of electromagnetic particles ($e^\pm$, $\gamma$) 
and muons ($\mu^{\pm}$) at ground. Most of EAS characteristics measured at ground 
are closely related to $X_{\max}$ and to $N_{\max}$~\cite{Schmidt:2007vq}. 
For example, the number of charged particles (mainly $e^{\pm}$) at ground $N_e$
strongly increases for deeper showers (where the maximal number of charged particles
is reached close to the ground) and falls down for decreasing $X_{\max}$ -- due to 
the quick shower attenuation in the atmosphere. On the other hand, the number of muons
$N_{\mu}$ at ground, which emerge mainly from charged pion and kaon decays,
has a weak dependence on $X_{\max}$, being instead sensitive to charged particle
multiplicity of hadron-air and nucleus-air collisions. This explains the importance
of the latter observable for CR composition studies with ground-based
detectors. Finally it should be noted that the number of muons is also
dependent on the physics of low-energy interactions in air showers. 
With pions typically decaying only after their energy is reduced
to $\sim$20~GeV, low-energy interactions of c.m.\ energy in the range
of $10-50$\,GeV become important~\cite{Drescher:2003gh,Meurer:2005dt}.\\

The relation between hadronic interactions at high energies and EAS observables 
has been studied numerically in detail in~\cite{Ulrich:2010rg}. Here we recall 
only the most important interaction parameters that determine the longitudinal development of 
air showers. The depth of shower maximum depends mainly on the characteristics 
of multiparticle production of the first few generations of hadronic interactions in a shower.
It is mainly related to the {\it inelastic cross section} of the primary incoming 
particle for interaction with air nuclei and on the corresponding
energy fraction transferred to secondary particles but the most energetic (``leading'') secondary
particle, relative to the primary particle, which is called {\it inelasticity}. Additionally, 
although less strongly, $X_{\max}$ depends on the inelastic cross sections and inelasticities
for interactions of secondary hadrons with air nuclei.
The third feature of direct relevance to the position of the shower maximum is the multiplicity
of the primary and subsequent very high-energy interactions, which defines how the energy is distributed
to secondary particles and corresponding sub-showers, i.e.\ whether many particles with small energy
or few particles with large energy are produced.\\

The LHC measurements of charged hadron pseudorapidity density and multiplicities
presented here, have considerable importance for CR physics. 
First, the observations by the ATLAS, CMS and ALICE collaborations indicate that $\dNdeta$ 
changes smoothly in the lab energy range from $10^{15}$ to $3\times 10^{16}$\,eV, 
and are well bracketed, at the $\pm 10$\% level at the highest energies, by the predictions 
of current interaction models used for EAS simulations. 
The bulk of hadron production data did not reveal serious deficiencies 
in the overall description of high energy hadronic collisions implemented in the event 
generators currently used in cosmic ray physics. 
This gives a strong support to the overall interpretation of the experimental results 
in terms of primary CR spectrum and nuclear mass composition in the knee energy range
($\unit[10^{15.5}]{eV}$) obtained by various collaborations that applied these models for 
the corresponding data analysis~\cite{Aglietta:2004np,Antoni:2005wq,Amenomori:2008jb,Budnev:2009mc}. 
At the same time, none of the models is in perfect agreement with all the hadronic observables
measured at the LHC (see Table~\ref{tab:rft_summary}), underlining the need for re-tuning model 
parameters and reconsidering model assumptions. New versions of 
\epos~\cite{Pierog:2010dt,Werner:2010ss} and \qgsjet~\cite{Ostapchenko:2010vb} 
are under development, partially influenced by the input from the LHC data.\\

Second, the new experimental results strongly disfavour speculative ideas that the knee in the 
observed cosmic ray spectrum may be due to a sudden change in the hadronic interaction mechanism 
above $2$\,TeV c.m.\ energy (see, for example,~\cite{Petkov:2005gv,Petrukhin:2006kp,Barcelo:2009uy,Dixit:2009mt}). 
Indeed, any generic model that has about 20\% of primary energy being transferred to 
particles that are not observed in EAS experiments just above $3\times 10^{15}$\,eV lab, would 
predict that proton showers of primary energy $3\times 10^{16}$\,eV, corresponding to about $7$\,TeV c.m.\
energy, would have to be reconstructed with a $25$\% lower energy in order to attribute the 
knee in the spectrum solely to the production of such new exotic particles.
Of course, so far only data from central LHC detectors have been published, covering just a
small fraction of the energy released in final state particles. However, it is difficult to imagine 
how such a significant change of the interaction properties from $2$ to $7$\,TeV c.m.\ energy
should only affect the particles close to the beam axis. More
realistically, one has to include also nuclei as cosmic rays in the knee
energy range, leading to even larger energy fractions that need to be
channeled to undetected particles in hadronic interactions.\\

Third, the first measurements at $\sqrt{s} = 7$\,TeV support a conventional extrapolation of the 
known features of multiparticle production to higher energies. Still it is not clear, whether the new data of
the Pierre Auger Collaboration~\cite{Abraham:2010yv}, indicating a
change to rather small shower-to-shower
fluctuations that would be typical for a primary mass composition
dominated by medium or heavy elements, could not also be attributed to 
a change of the characteristics of hadronic interactions at energies
above $3\times 10^{18}$\,eV~\cite{Wibig:2009zza,Wibig:2009zz}.
But the current wide range of predictions for the particle densities,
$\dNdeta\approx$~10 (\epos, \sibyll)~--~50~(\qgsjet II), 
as well as for the mean hadron transverse momentum, 
$\meanpt\approx$~0.6~(\sibyll, \qgsjet 01)~--~1~(\epos)~GeV/c,
justifies the concurrent use of various MCs to gauge the uncertainties connected 
to hadronic interaction models in the interpretation of the cosmic ray
data at GZK-cutoff energies.\\

Last but not least, the measurements of $\pT$-differential spectra and 
mean $\pT$ of hadrons at LHC midrapidity do not have a direct
impact on the interpretation of air shower data because the shape of
lateral distributions of electrons (positrons) and muons at ground
level is rather defined by multiple Coulomb scattering in the
atmosphere and by the transverse momentum spectra of secondaries at much
lower (fixed-target) energies~\cite{Drescher:2003gh,Meurer:2005dt}.
Nevertheless, the corresponding results are of importance for testing
the overall physics consistency of 
soft and hard interaction mechanisms implemented in the models as well as for the
understanding of the parton saturation and multi-scattering phenomena discussed above.

\subsection{Importance of forthcoming LHC measurements}

The first multiplicity measurements at $\sqrt{s} = 7$\,TeV ($2.5
\times 10^{16}$\,eV in lab system) have put already serious constraints on
the interpretation of cosmic ray data. Coming \pp\ data at the nominal c.m.~energy of 
$\sqrt{s} = 14$\,TeV, corresponding to cosmic-ray protons of $10^{17}$\,eV in the lab frame,
will further reduce the uncertainty linked to the 
extrapolation of the interaction models to ultra-high energies, as they will
the expected proton-nucleus runs ($p Pb$ collisions at $\sqrtsnn$~=~8.8~TeV)~\cite{dEnterria:2008jk,pA}.
First results on the $\dNdeta$, and its centrality dependence, measured in Pb-Pb collisions 
at $\sqrtsnn$~=~2.76~TeV~\cite{Collaboration:2010cz} 
will provide extra important cross-checks on the role of initial-state
gluon saturation effects in the hadronic wave functions.\\

There are many more measurements that can be done already at the
current LHC energy to reduce the uncertainties of air shower
predictions. Key information is expected, for example, from the measurement 
of the distribution of forward neutral hadrons ($\pi^0$ and neutrons) 
by LHCf~\cite{lhcf} and possibly other experiments using zero degree 
and other forward calorimeters~\cite{Grachov:2006ke,White:2010zzd,Oppedisano:2009zz}
in both \pp\ and Pb-Pb interactions. 
Of similar importance would be the measurement of the energy flow and particle 
spectra in the very forward direction, in the pseudorapidity range
from 5 to 10. This is an angular range that is very difficult to
access in collider experiments but it is partially covered by
various detectors in the LHC experiments~\cite{d'Enterria:2007dt},
for example, TOTEM~\cite{Anelli:2008zza} and CASTOR (CMS)~\cite{Andreev:2010zzb}.
Another fundamental input for tuning and extrapolating models are the
total, elastic, and diffractive cross sections accessible to measurement 
thanks to various forward proton  detectors (existing such as TOTEM, ATLAS 
ALFA~\cite{Ask:2007fr}, or proposed~\cite{Albrow:2008pn}) in the LHC tunnel area. 
Examples of the impact of the current uncertainty of the cross
section on air shower observables are discussed in~\cite{Ulrich:2009yq}.\\

\begin{figure}[htbp]
\includegraphics[width=16.cm,angle=0]{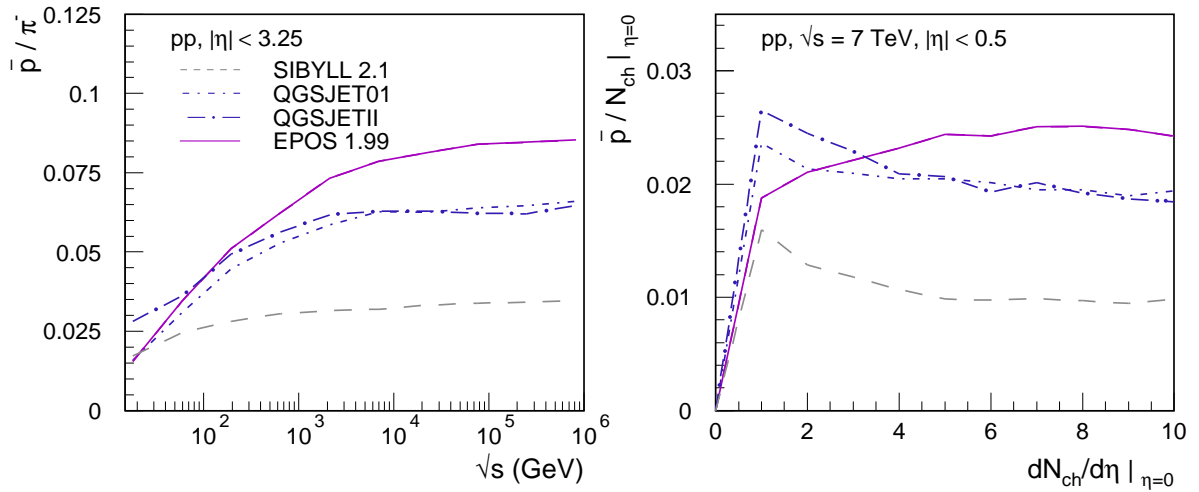}
\hfill
\hspace*{7cm}
\caption{Examples of midrapidity distributions whose measurement in \pp\ collisions at $\sqrt{s} = 7$\,TeV
will help to refine the modeling of (anti)baryon production in current event generators. 
Left: Collision-energy dependence of the ratio of $\bar{p}$ to $\pi^-$
multiplicities. Right: Ratio of  $\bar{p}$ to charged particle multiplicities 
as a function of the central pseudorapidity density.
}
\label{fig:ap-production}
\end{figure}

It is worth mentioning that other measurements that can be performed with
central detectors at the LHC, such as e.g.~of the rapidity spectra of identified hadrons, 
are of significant importance for cosmic ray physics. Indeed, due to different lifetimes of charged
pions and kaons, knowledge of their relative contributions to the 
multiplicity of secondaries is required for precise calculations of 
the total number and of the energy spectra of muons produced during
EAS development. Also the production of antibaryons is a theoretically poorly
understood aspect of hadronic interactions and data on baryon pair
production are sparse.  As demonstrated in~\cite{Pierog:2006qv}, the
muon content of air showers may be significantly enhanced if the rate
of production of (anti)baryons is up to $30$\% higher than assumed in
models like \qgsjet\ and \sibyll. Baryon-induced subshowers lead to a
higher number of muons at ground than meson-induced ones.  The energy
dependence of the overall baryon production rate, its relation to
the centrality of the collision, and the momentum distribution
of the baryons are important quantities to be measured.
This is illustrated in Fig.~\ref{fig:ap-production} where CR model predictions
for antiproton production in \pp\ at $\sqrts$~=~7~TeV are shown. The number of
fragmenting strings leads to a so-called delayed energy threshold for
baryon pair production. This threshold is lowest in \sibyll, being
well below the Tevatron energy, and largest in \epos\ where it is in the
energy range of the LHC (Fig.~\ref{fig:ap-production}, left). Also their
predicted asymptotic limits are very different. The dependence of 
antiproton production on the midrapidity particle density in proton-proton 
collisions at the LHC is shown in Fig.~\ref{fig:ap-production}~(right). 
The peak at low central multiplicities is related to the fact that antibaryons are
pair-produced with baryons in a single string, consuming a large
fraction of the energy stored in the corresponding string. Peripheral
interactions are typically simulated with only two color-strings
maximizing the energy per string. \epos\ differs not only
quantitatively but also qualitatively in the antibaryon-baryon pair
production model from the other models~\cite{Liu:2003wja}.



%
%

\section{Summary}
\label{sec:summ}

We have compared first CERN LHC data on inclusive charged hadron production -- central 
pseudorapidity densities $\dNdeta$, multiplicity probabilities $P(N_{ch})$, and mean transverse 
momentum $\meanpt$ -- with hadronic interaction models (\qgsjet 01 and II, \sibyll, and \epos~1.99)
used for the simulation of extensive air showers generated by ultra-high-energy cosmic rays in the Earth 
atmosphere. We have also included in this comparison commonly used collider physics 
event generators, such as \pythia\ (various tunes of version 6.4, and the new \pythia\ 8)
and \phojet, in order to identify the model ingredients most sensitive to bulk hadron 
production. In all the cases we considered only those versions of the interaction models
that were developed and tuned before the LHC data became available.\\

In contrast to most Monte Carlo event generators for collider physics, models 
developed for air shower simulation are based on Reggeon Field Theory (RFT)
and are optimized to predict the overall characteristics of the final 
state particles including those of predominantly soft interactions. Moreover,
these models are designed to deal with proton-nucleus and nucleus-nucleus 
interactions and allow for an extrapolation to energies as high as
$\sqrt{s} \sim 400$\,TeV. Therefore the comparison to LHC data is an
important benchmark for the quality of the models and hence the
reliability of air shower simulations currently used to interpret
cosmic ray (CR) data.\\ 

The quality of the LHC data description varies from model to model and differs 
for different observables. A first general observation is that none of the 
models considered provides a very good description of all the LHC data considered here. 
Yet, the CR models bracket the LHC central rapidity densities and the low region
of the multiplicity probabilities more naturally than most of the \pythia\ tunes. 
The main difference appears to be on the more advanced treatment of diffractive 
scattering in the RFT models compared to the latter.
The measured midrapidity charged-particle density is found to follow a 
simple power-law in energy, $s^\epsilon$, from 10 GeV up to 7~TeV c.m.~energy 
with exponent $\epsilon\approx$~0.10. This observation can be used to retune some 
basic model ingredients and improve their extrapolations to the highest energies. 
Such a result constrains in particular the way in which multiparton interactions 
and gluon saturation are implemented in various MCs (e.g.~in \pythia, \phojet\ 
and \sibyll), via an energy-dependent infrared transverse momentum cutoff, 
$Q_0(\sqrts)$, for (multi)parton scatterings.\\

The studies presented here have also shown that it is very difficult, 
if not impossible, to compare model predictions with data
at the sub-$5$\% level if the trigger used for data-taking cannot be
implemented without a full detector simulation. It is highly
desirable to have the LHC inelastic, minimum bias and diffractive 
data published with a simple hadron-level prescription for emulating 
the data trigger, to which the measurements have been corrected to. 
Only then we will be able to make best use of LHC results to lower the 
uncertainty of various non-perturbative model ingredients and to 
improve the description of the transition from soft to hard parton dynamics 
(multiparton interactions, gluon saturation, ...).\\

With $\sqrt{s} = 7$\,TeV, LHC is the first collider reaching an
energy higher than the knee in the energy spectrum of cosmic rays. 
The LHC hadron multiplicity measurements give strong support to the conventional
interpretation that 
the break in the power-law index of the observed CR spectrum at $\unit[10^{15.5}]{eV}$ 
is indeed due to a feature of the primary cosmic ray flux. Alternative interpretations of the
knee being a side-effect of rapidly changing properties of hadronic interactions 
above $\sqrt{s} \approx 2$\,TeV are strongly disfavoured. Similarly the LHC
measurements support the interpretation of air shower data in the knee energy 
range as reflecting a change from a light to a more heavy mass composition.
No new or exotic physics assumptions or extrapolations are needed for describing 
the overall event features measured in the central pseudorapidity region at the LHC. 
While re-tuning of model parameters to match LHC data will improve the reliability 
of air shower simulations, there is no indication from the LHC results that the
extrapolations have to be changed significantly. At the highest CR energies 
of $\mathcal{O}$(10$^{20}$~eV) -- i.e.~more than twenty times higher than those 
c.m.~energies reachable in \pp\ at the LHC -- 
the current wide range of predictions for the particle densities,
$\dNdeta\approx$~10 (\epos, \sibyll)~--~50~(\qgsjet II), 
as well as for the mean hadron transverse momentum, 
$\meanpt\approx$~0.6~(\sibyll, \qgsjet 01)~--~1~(\epos)~GeV/c,
justify today the concurrent use of all RFT MCs to gauge the uncertainties linked 
to the underlying hadronic interactions in the interpretation of the cosmic ray
data at GZK-cutoff energies.\\

Future measurements at the LHC will further cross-check interaction models
and help to understand better the underlying hadron production processes. 
Of direct relevance to the interpretation of air shower data are the 
multiplicity and spectrum of produced high-energy particles
(leading particles, neutral and charged) emitted at forward rapidities, 
the contribution of different particle types to the overall final state 
multiplicity, the inelastic cross section and its breakdown in various 
diffractive contributions, as well as the production of baryon-antibaryon 
pairs at mid-rapidity in proton-proton, nucleus-nucleus and proton-nucleus collisions.\\

\section*{Acknowledgments} 

We thank Ferenc Sikler and Gabor Veres (CMS) and Jan-Fiete Grosse-Oetringhaus (ALICE)
for valuable discussions on the LHC experimental data. Enlightening discussions with 
Torbj\"orn Sj\"ostrand on \pythia\ 6 and 8 are also acknowledged. This work has been supported in 
part by Bundesministerium f{\"u}r Bildung und Forschung (BMBF) grant No.~05A08VK1. 
D.d'E. acknowledges support by the 7th EU Framework Programme (contract FP7-ERG-2008-235071). 
S.O. acknowledges the support of the FP7 Marie Curie program (contract FP7-IEF-2007-220251) 
and by the program Romforskning of Norsk Forskningsradet.



\end{document}